\documentclass[acmsmall,authorversion,nonacm]{acmart}

\usepackage{listings}
\usepackage{graphicx}
\usepackage{dsfont}
\usepackage{amsthm}
\usepackage{amsmath}
\usepackage{algorithm}
\usepackage{algpseudocode}
\usepackage{makecell}
\usepackage{caption}
\usepackage{subcaption}
\theoremstyle{definition}
\newtheorem{example}{Example}[section]
\usepackage[braket, qm]{qcircuit}

\DeclareMathOperator*{\argmin}{arg\,min}

\algrenewcommand\algorithmicrequire{\textbf{Input:}}
\algrenewcommand\algorithmicensure{\textbf{Output:}}

\setcopyright{acmlicensed}
\copyrightyear{2018}
\acmYear{2018}
\acmDOI{XXXXXXX.XXXXXXX}

\acmJournal{JACM}
\acmVolume{37}
\acmNumber{4}
\acmArticle{111}
\acmMonth{8}

\begin{document}

\title{Left-Deep Join Order Selection with Higher-Order Unconstrained Binary Optimization on Quantum Computers}

\author{Valter Uotila}
\email{valter.uotila@helsinki.fi}
\orcid{1234-5678-9012}
\affiliation{%
  \institution{University of Helsinki}
  \country{Finland}
}

\renewcommand{\shortauthors}{Uotila}

\begin{abstract}
Join order optimization is among the most crucial query optimization problems, and its central position is also evident in the new research field where quantum computing is applied to database optimization and data management. In the field, join order optimization is the most studied database problem, usually tackled with a quadratic unconstrained binary optimization model, which is solved with various meta-heuristics such as quantum annealing, quantum approximate optimization algorithm, or variational quantum eigensolver. In this work, we continue developing quantum computing techniques for join order optimization by presenting three novel quantum optimization algorithms. These algorithms are based on a higher-order unconstrained binary optimization model, which is a generalization of the quadratic model and has not previously been applied to database problems. Theoretically, these optimization problems naturally map to universal quantum computers and quantum annealers. Compared to previous research, two of our algorithms are the first quantum algorithms to precisely model the join order cost function. We prove theoretical bounds by showing that these two methods encode the same plans as the dynamic programming algorithm without cross-products, which provides the optimal result up to cross-products. The third algorithm reaches at least as good plans as the greedy algorithm without cross-products. These results set an important theoretical connection between the classical and quantum algorithms for join order selection, which has not been studied in the previous research. To demonstrate our algorithms' practical usability, we have conducted an experimental evaluation on thousands of clique, cycle, star, tree, and chain query graphs using quantum and classical solvers.
\end{abstract}





\maketitle

\section{Introduction}

Join order optimization is one of the critical stages in query optimization where the goal is to determine the most efficient sequence in which joins should be performed \cite{10.1145/582095.582099}. The join order can significantly affect query performance, especially in large databases \cite{Neumann_Radke_2018}. Join order optimization is a well-researched NP-hard problem \cite{10.1145/1270.1498} with various exhaustive and heuristic solutions \cite{10.14778/2850583.2850594, 10.1007/s007780050040}.

The central position of join order optimization in database research is also evident in the new research field where quantum computing is applied to database optimization and data management \cite{Schonberger_2022, Uotila_2022}. In this subfield, the join order selection problem is the most studied \cite{Schonberger_Scherzinger_Mauerer, Winker_Calikyilmaz_Gruenwald_Groppe_2023, Schonberger_Trummer_Mauerer_2023, Nayak_Winker_Groppe_Groppe_2024, Franz_Winker_Groppe_Mauerer_2024,10.14778/3632093.3632112, DBLP:conf/q-data/SaxenaSS24}. The other quantum computing for database and data management-related problems includes index selection \cite{Gruenwald_Winker_Groppe_Groppe_2023, DBLP:conf/q-data/TrummerV24}, cardinality and metric estimations \cite{Uotila_2023_sqlcircuits,DBLP:conf/q-data/KittelmannSS24}, transaction scheduling \cite{Bittner_Groppe_2020b}, resource allocation \cite{Uotila_Lu_2023}, schema matching \cite{Fritsch_Scherzinger_2023} and multiple query optimization \cite{Trummer_Koch_2016}.

Gaining an advantage of quantum algorithms over classical algorithms has proved extremely challenging in real-life applications, including the listed database applications. Depending on the definition of quantum advantage, some famous experiments \cite{Arute_Arya_Babbush_Bacon_Bardin_Barends_Biswas_Boixo_Brandao_Buell_etal2019,Harrow_Montanaro_2017,Kim_Eddins_Anand_Wei_vandenBerg_Rosenblatt_Nayfeh_Wu_Zaletel_Temme_etal_2023,King_Nocera_Rams_Dziarmaga_Wiersema_Bernoudy_Raymond_Kaushal_Heinsdorf_Harris_etal_2024,Madsen_Laudenbach_Askarani_Rortais_Vincent_Bulmer_Miatto_Neuhaus_Helt_Collins_etal_2022,Zhong_Wang_Deng_Chen_Peng_Luo_Qin_Wu_Ding_Hu_etal_2020, Zhu_Cao_Chen_Chen_Chen_Chung_Deng_Du_Fan_Gong_etal_2021} demonstrate specific advantages, but they are not known to have any real-life applications. Thus, no real-life application of quantum computing is widely accepted to demonstrate quantum advantage in any field. On the other hand, some algorithms, such as Shor's and Grover's algorithms, show a provable advantage over the best classical algorithm on fault-tolerant quantum computers, which do not yet exist. These theoretical results are a key motivation for developing better quantum computing hardware, and this work aims to contribute to this algorithmic development.

The difficulty of applying quantum computing lies in the fundamentally different computational models~\cite{Nielsen_Chuang_2010}, different complexity analyses \cite{aaronson2016complexitytheoreticfoundationsquantumsupremacy}, the probabilistic nature of quantum computing, small-scale and erroneous quantum hardware, barren-plateaus in training and optimization landscapes \cite{Ragone_Bakalov_Sauvage_Kemper_Ortiz_Marrero_Larocca_Cerezo_2024} and the fact that it seems complicated to develop high-performing quantum algorithms. The last property is supported by the fact that the most essential quantum algorithm primitives can be listed on a single webpage \cite{quantum_algorithm_zoo}. 

Since showing that the current quantum computers provide any benefit has been challenging, \cite{10.14778/3632093.3632112} suggested moving from quantum hardware to quantum-inspired hardware, especially in join order optimization. They argued that we should study special-purpose solvers and hardware, including digital annealers. While they showed that this direction is promising, they did not extensively examine the possible benefits of modifying the underlying quantum optimization model, which has been similar to the MILP solution \cite{Trummer_Koch_2017}.

As a continuation of the idea to revise the underlying assumptions about hardware (relaxing from quantum to quantum-inspired), we suggest applying a special optimization model, a higher-order binary optimization model, which is a relaxation of the previously widely used quadratic model \cite{Schonberger_Scherzinger_Mauerer,Trummer_Koch_2017,10.14778/3632093.3632112,Schonberger_Trummer_Mauerer_2023,Nayak_Winker_Groppe_Groppe_2024, DBLP:conf/q-data/SaxenaSS24}. If we seek (database) applications that are likely to benefit from quantum computing, one of our central arguments is that we might want to move from quadratic to higher-order models. Focusing on the quantum computing paradigm that is restricted to quadratic interactions between qubits, the research has shown that there are only particular problems where these devices beat classical computers \cite{PhysRevX.6.031015,King_Nocera_Rams_Dziarmaga_Wiersema_Bernoudy_Raymond_Kaushal_Heinsdorf_Harris_etal_2024}. On the other hand, there is no evidence that this advantage would transfer to practically relevant problems \cite{Willsch_Willsch_Gonzalez_Calaza_Jin_De_Raedt_Svensson_Michielsen_2022}. One of the key challenges is ''quadratization'', which requires that the real-life problem is translated into a quadratic format. Translating a practical problem into this format is often so expensive that we lose the potential advantage, even in theory. For example, the costly problem encoding is evident in \cite{Nayak_Winker_Groppe_Groppe_2024} where the number of binary variables grows exponentially in terms of relations in a database.

Additionally, the previous quantum computing formulations for the join order selection problem did not benefit from the query graph's structure, which we encode in the optimization model. By using information from the query graphs and assuming that the cross products are expensive, we can decrease the size of the optimization problems. The other critical scalability finding lies in the selection of binary variables. With a clever choice of binary variables, we can compute the cost precisely and reduce the number of variables and their types. The previous research \cite{Schonberger_Scherzinger_Mauerer, 10.14778/3632093.3632112} has used four variable types (variables for relations, joins, predicates, and cost approximation). We decrease this number to one variable type, which works in most cases except for clique graphs, which require two types.

Quantum computing research for database applications has not provided many theoretical results about the performance of their methods. In this work, we prove two bounds for our methods, which connect the quantum algorithms to the classical ones. This is important because it helps us comprehend the capabilities of the current quantum computation solutions compared to the established classical methods.


The key contributions are as follows:
\begin{enumerate}
    \item We develop three novel higher-order unconstrained binary optimization algorithms to solve the join order selection problem on universal quantum computers and quantum annealers.
    \item We provide theoretical bounds that characterize the accuracy of our algorithms.
    \item We perform a comprehensive experimental evaluation with varying quantum and classical solvers demonstrating the proposed algorithms' practical usability within the limits of the current hardware.
\end{enumerate}

The structure of the paper is as follows. First, we formally state the join order selection problem, discuss the quadratic and higher-order binary optimization models, and define their connection to quantum computing. Then, we present the main algorithms. We prove the theoretical bounds for the accuracy of the methods. We summarize the results from the experimental evaluation and discuss how our contributions relate to the previous research solving the join order selection problem with quantum computing. On GitHub \cite{anonymous2024qjoin} and in the experimental results, we refer to the implementation of our algorithms as Q-Join.
\section{Setting}

\subsection{Join order selection problem}

\begin{figure}[tb]
    \centering
    \includegraphics[width = 0.3\columnwidth]{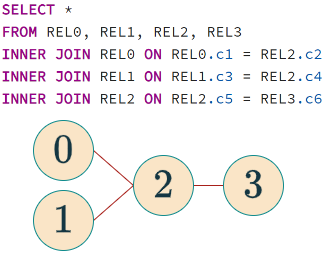}
    \caption{SQL query and its corresponding query graph}
    \label{fig:SQL_query_graph}
    \Description[SQL query and its corresponding query graph]{SQL query is represented as text containing the standard SQL keywords and four inner joins between five tables. This query corresponds to a tree-shaped query graph, and its corresponding query graph contains five nodes (tables). The inner join conditions in the query give the edges.}
\end{figure}

We start by formally defining the join order selection problem. We assume the SQL queries are given as query graphs of form $G = (V, E)$, where $V$ is the set of nodes (i.e., tables or relations). Fig.~\ref{fig:SQL_query_graph} shows an example of a query and its query graph. We denote relations as $R_i$ for some non-negative integer $i$. The set $E$ is the set of edges defined by join predicates $p_{ij}$ between tables $R_i$ and $R_j$. Every table $R_i$ has a cardinality, denoted by $|R_i|$, and every predicate has a selectivity $0 < f_{ij} \leq 1$. The join is denoted by $R_i \bowtie_{p_{ij}} R_j$. This work assumes that the joins are inner joins, although we will discuss the extension to other joins, such as outer joins.

A join tree $T$ of a query graph $G$ is a binary tree where every relation of $G$ appears once in the leaf nodes, and every non-leaf node has a form $R_{k_1} \bowtie \ldots \bowtie R_{k_n}$ for some $R_{k_1}, \ldots, R_{k_n} \in G$ such that the two child nodes of $R_{k_1} \bowtie \ldots \bowtie R_{k_n}$ contain the relations $R_{k_1}, \ldots, R_{k_n}$. Following the definition in \cite{Neumann_Radke_2018}, we define that a join tree $T$ adheres a query graph $G$ if for every subtree $T' = T_1 \bowtie T_2$ of $T$ there exists relations $R_1$ and $R_2$ such that $R_1 \in T_1$ and $R_2 \in T_2$ and $(R_1, R_2) \in E$. The join order selection problem is finding a join tree $T$ that adheres to query graph $G$ and minimizes a given cost function. In the terminology of this paper, the first problem, a join tree adhering to the query graph, is called a \textit{validity constraint}. The problem of minimizing the cost is called a \textit{cost constraint}. Next, we define the standard cost function for join trees.

Standard cost functions are based on estimating cardinalities of intermediate results in the join order process \cite{10.14778/2850583.2850594, Neumann_course, cluet1995complexity}. Thus, we first define how to compute the cardinality of a given join tree $T$. For join tree $T$, its cardinality is defined recursively
\begin{equation}\label{eq:join_tree_card}
    |T| = \begin{cases}
    \ |R_i| &\quad \text{if } T = R_i \text{ is a leaf node} \\
    \ \prod_{R_i \in T_1, R_j \in T_2} f_{ij}|T_1||T_2| &\quad \text{if } T = T_1 \bowtie T_2.
    \end{cases}
\end{equation}

Based on the cardinalities, we define the cost function recursively as
\begin{equation}\label{eq:cost_function}
    C(T) = \begin{cases}
        \ 0 &\quad \text{if } T = R_i \text{ is a leaf node} \\
        \ |T| + C(T_1) + C(T_2) &\quad \text{if } T = T_1 \bowtie T_2.
    \end{cases}
\end{equation}
This is the standard cost function \cite{cluet1995complexity}, which has also been used in earlier quantum computing formulations \cite{Schonberger_Scherzinger_Mauerer, 10.14778/3632093.3632112, Schonberger_Trummer_Mauerer_2023} and in the corresponding MILP formulation \cite{Trummer_Koch_2017}.

\subsection{Unconstrained binary optimization}

Optimization is one of the key fields where quantum computing is assumed to provide computational value \cite{abbas2024quantumoptimizationpotentialchallenges}. This part provides a brief and high-level overview of how optimization algorithms are developed in quantum computing. We guide a reader to \cite{Nielsen_Chuang_2010, Winker_Groppe_Uotila_Yan_Lu_Franz_Mauerer_2023} to more detailed basics about quantum computing. Additionally, \cite{Schonberger_Scherzinger_Mauerer} provides an excellent introduction to quantum annealing and quadratic unconstrained binary optimization models for a database specialist.

Our work relies on Higher-order Unconstrained Binary Optimization (HUBO) \cite{boros_2002} problems, which are a generalization of Quadratic Unconstrained Binary Optimization (QUBO) problems. As far as we know, there is little research on formulating domain-specific problems using HUBOs. One use case for HUBOs has been optimizing matrix multiplication algorithms \cite{uotila2024tensordecompositionsadiabaticquantum}. One reason for this is that HUBO problems are hard not only theoretically but also practically \cite{boros_2002}. Due to this computational complexity, they provide a potential area for exploring the practical benefits of quantum computing over classical approaches. Despite being challenging, they have a straightforward quantum computational formulation \cite{10313783} in theory.

Next, we define HUBO problems \cite{boros_2002} formally and show their connection to QUBO problems. Let $x \in \left\{0, 1 \right\}^{n}$ be a binary variable vector of type $x = (x_1, \ldots, x_n)$ so that $x_i \in \left\{ 0, 1 \right\}$ representing values false and true. Let $[n] = \left\{1, \ldots, n \right\}$ be an indexing set. The HUBO problem is the following minimization problem of a binary polynomial
\begin{equation}\label{eq:hubo}
    \argmin_{x \in \left\{0, 1 \right\}^{n}} \ \sum_{S \subset [n]}\alpha_{S}\prod_{i \in S}x_i,
\end{equation}
where $\alpha_{S} \in \mathds{R}$. For each non-empty subset $S$, we have the corresponding higher-order term $\alpha_{S}\prod_{i \in S}x_i$. In practice, we have $\alpha_{S} = 0$ for many terms. Otherwise, in the worst case, we have $2^{|S|} - 1$ terms. In this work, \textit{term} will sometimes mean the variable-tuple $\prod_{i \in S}x_i$ (excluding the coefficient $\alpha_{S}$), but in those cases, we will explicitly indicate what the coefficient is. Alternatively, we can write the same polynomial as
\begin{equation}\label{def:hubo_open}
    \sum_{S \subset [n]}\alpha_{S}\prod_{i \in S}x_i = \sum_{i \in [n]}\alpha_i x_i + \sum_{i < j}\alpha_{i,j}x_ix_j + \sum_{i < j < k}\alpha_{i,j,k}x_ix_jx_k + \ldots
\end{equation}

Quadratic Unconstrained Binary Optimization (QUBO) problems are a restricted case of HUBO problems where we consider only limited-sized subsets $|S| \leq 2$. Concretely, a QUBO problem is the minimization problem of the polynomial
\begin{equation}\label{def:qubo}
    \sum_{i \in [n]}\alpha_i x_i + \sum_{i < j}\alpha_{i,j}x_ix_j.
\end{equation}
Both QUBO and HUBO problems are NP-hard \cite{lucas_2014, boros_2002}. As discussed, QUBO formalism has been the standard method for tackling database optimization problems, and many other well-known optimization problems (e.g., knapsack, maxcut, and traveling salesman) have a QUBO formulation \cite{lucas_2014}.

\subsection{Optimization on quantum hardware}

Next, we briefly introduce the basics of quantum computing for optimization problems and discuss techniques for solving HUBOs and QUBOs on quantum hardware. Quantum computing can be divided into multiple paradigms regarding hardware and software. This division is exceptionally versatile since there is no "winning" method for building quantum computers yet. Quantum computers are designed to be built on superconducting circuits (IBM, Google, IQM) \cite{Wendin_2017}, trapped ions (Quantinuum, IonQ) \cite{paul1990electromagnetic}, neutral atoms (Quera, Pasqal) \cite{GRIMM200095}, photons (Xanadu) \cite{knill2001scheme}, diamonds (Quantum Brilliance) \cite{neumann2008multipartite}, and many other quantum mechanical phenomena \cite{https://doi.org/10.1002/spe.3039}. A special type of quantum hardware is a quantum annealer (D-Wave) \cite{APOLLONI1989233, PhysRevE.58.5355}, which does not implement universal quantum computation but offers specific optimization capabilities with better scalability.

On top of the hardware, a partly hardware-dependent software stack is designed to translate and compile high-level quantum algorithms into a format that specific quantum hardware supports. As we will explain, QUBOs are a widely accepted high-level abstraction that can be solved on most quantum hardware. The other common high-level abstraction is quantum circuits. Quantum algorithm design can still be divided into paradigms: adiabatic and circuit-based quantum computing, which are universal quantum computing paradigms \cite{Aharonov_van_Dam_Kempe_Landau_Lloyd_Regev_2004}. Unlike traditional introductions to quantum computing, we focus on the fundamentals of adiabatic quantum computing. This choice is motivated by the fact that our experimental evaluation was conducted on a quantum annealer, a type of adiabatic quantum computer. Due to space limitations, we introduce the more commonly used quantum circuit model in the appendix. With the circuit model, we have another set of tools to optimize unconstrained binary optimization problems. Since the scalability of universal quantum computers using quantum circuits is still very limited, we present the connection between our algorithms and the quantum circuit model only theoretically and hope that our work is a motivating and practical use case.

\textbf{Adiabatic quantum computing.} Quantum computing can be implemented utilizing the adiabatic evolution of a quantum mechanical system \cite{Farhi_Goldstone_Gutmann_Sipser_2000}. This work utilizes quantum annealing, which is a part of the adiabatic quantum computing paradigm \cite{Lidar_adiabatic}. We start from the axiomatic fact that the Schrödinger equation describes an evolution of the quantum mechanical system \cite{Nielsen_Chuang_2010}. This evolution models a system with a Hermitian operator known as a Hamiltonian. In this work, we are not interested in arbitrary Hamiltonians but in those with a form that maps to QUBO and HUBO optimization problems. For QUBOs, the corresponding problem Hamiltonian, also called an Ising Hamiltonian, is
\begin{displaymath}
\sum_{i} h_i \sigma_{z}^{i} + \sum_{j < i} J_{i,j}\sigma_{z}^{i}\sigma_{z}^{j},
\end{displaymath}
where $\sigma_{z}^{j}$ are Pauli-Z operators for each $j$. The correspondence between the formulation in Eq. \eqref{def:qubo} is clear: the coefficients $h_i$ are the linear terms, and $J_{i,j}$ are the quadratic terms. For each $i$, $\sigma_{z}^{i}$ corresponds to the variable $x_i$. For HUBOs the problem Hamiltonian is
\begin{displaymath}
    \sum_{S \subset [n]}\beta_{S}\prod_{i \in S}\sigma_{z}^{i},
\end{displaymath}
where the correspondence to Eq. \eqref{def:hubo_open} is the same. As in the case of QUBOs and HUBOs, we aim to minimize the value of a Hamiltonian. In other words, we aim to find a quantum state, called a ground state, where the Hamiltonian's energy is minimized. After minimizing the Hamiltonian, we also obtain a solution to the corresponding combinatorial optimization problem \cite{Farhi_Goldstone_Gutmann_Sipser_2000}.

Since a Hamiltonian $H$ is a Hermitian operator by definition, it has the following spectral decomposition
\begin{displaymath}
    H = \sum_{e} \lambda_{e}|e\rangle \langle e|,
\end{displaymath}
where $\lambda_{e}$ are the operator's eigenvalues and $|e\rangle$ are its eigenstates. Minimizing the Hamiltonian and solving the corresponding combinatorial optimization problem require finding the lowest eigenvalue $\lambda_{e}$ and its corresponding eigenstate, also called a ground state. This is the central problem addressed by the quantum computing methods introduced in this work.

Next, we describe how adiabatic quantum computing can find the lowest eigenstate and solve the optimization problem. For simplicity, let us focus on solving QUBOs on a quantum annealer. The general adiabatic quantum computing is similar \cite{Lidar_adiabatic}. We define the following Hamiltonian \cite{d_wave}
\begin{equation}\label{eq:adiabatic_Hamiltonian}
H(s) = \underbrace{-\frac{A(s)}{2}\sum_{i} \sigma_{x}^{i}}_{\text{initial Hamiltonian}} + \underbrace{\frac{B(s)}{2}\left( \sum_{i} h_i \sigma_{z}^{i} + \sum_{j < i} J_{i,j}\sigma_{z}^{i}\sigma_{z}^{j}\right)}_{\text{problem Hamiltonian}},
\end{equation}
where $A(s)$ is the so-called tunneling energy function and $B(s)$ is the problem Hamiltonian energy function at $s$. During the annealing process, the value $s$ runs from $0$ to $1$ , causing $A(s) \to 0$ and $B(s) \to 1$. We begin with a simple initial Hamiltonian, whose ground state is easy to prepare, and gradually evolve to the problem Hamiltonian. By the adiabatic theorem \cite{Lidar_adiabatic}, if the process is slow enough, the system remains in its ground state and ends up solving the optimization problem. However, quantum annealers are limited to solving QUBOs, meaning we cannot encode higher-order terms in the problem Hamiltonian. In contrast, universal adiabatic and gate-based quantum computers do not have this restriction.

\textbf{Classical computers.} Considering classical computers, we can solve QUBOs using simulated annealing \cite{doi:10.1126/science.220.4598.671}, digital annealing \cite{Aramon_Rosenberg_Valiante_Miyazawa_Tamura_Katzgraber_2019}, and classical solvers such as Gurobi and CPLEX. Unfortunately, classical solvers cannot natively solve higher-order binary optimization models, but we have to rely on rewriting methods that reduce HUBOs into QUBOs. We introduce two reduction methods to rewrite higher-order problems into quadratic ones. Due to space limitations, the details of these reduction methods are in the appendix. The key idea is to replace higher-order terms with slack variables.
\section{Join order cost as HUBO}

In this section, we develop two higher-order unconstrained binary optimization (HUBO) problems that encode the cost function in Eq.~\eqref{eq:cost_function} for a left-deep join order selection problem. Compared to the previous quantum computing for join order optimization research, we formulate the optimization problem from the perspective of \textit{joins} instead of the perspective of \textit{relations} \cite{Schonberger_Scherzinger_Mauerer}. Given a query graph $G$, the number of required joins to create a valid left-deep join tree is $|V| - 1$, where $|V|$ is the number of nodes (i.e., relations, tables) in query graph $G$. This is easy to see since the first join is performed between two tables, and after that, every join includes one more table until all the tables have been joined.

Our algorithm is designed to rank joins, and ranking gives the order for the joins. This means a join (i.e., edge in the query graph $G$) has a rank $0 \leq k < |V| - 1$ if the join should be performed after all the lower rank joins are performed. We need $|V| - 1$ rank values to create a left-deep join order plan. Having $|V| - 1$ rank values applies to left-deep join plans but not bushy ones. For bushy plans, we can join multiple tables simultaneously, meaning that some of the joins can have the same rank, i.e., appear at the same level in the join tree.

Initially, any join $(R_i, R_j) \in G$ can have any rank $0 \leq r < |V| - 1$. We define the binary variables of our HUBO problems to be
\begin{equation}\label{def:binary_variables}
    x_{i,j}^{r} \in \left\{0, 1 \right\},
\end{equation}
where the indices $i$ and $j$ refer to the relations $R_i$ and $R_j$ and $r$ denotes the rank. Hence, our model consists of $(|V|- 1)|E|$ binary variables since for every rank value $0 \leq r < |V| - 1$, we have $|E|$ many joins (edges) from which we can choose the join.

The interpretation of these binary variables is as follows: If $x_{i,j}^{r} = 1$, then the join $(R_i, R_j)$ should be performed at rank $r$. Now the join $(R_i, R_j)$ is not necessarily between the tables $R_i$ and $R_j$ since at $r > 0$ the left relation is an intermediate result of type $R_{k_1} \bowtie \ldots \bowtie R_i \bowtie \ldots \bowtie R_{k_n}$ for some indices $k_1, \ldots, k_n$. Thus, the tuples $(R_i, R_j)$ represent joins in the query graph rather than materialized joins in query processing.

\begin{example}
Consider that we have a simple, complete query graph of four relations $\left\{0,1,2,3\right\}$. Thus, we have $|V| = 4$ relations, $|E| = 6$ possible joins and $(|V| - 1)|E| = 18$ binary variables. Depending on the selectivities and cardinalities, an example solution that the model can return is $x_{0,1}^{0} = 1$, $x_{1,2}^{1} = 1$, and $x_{2,3}^{2} = 1$, which gives us left-deep join tree $[[0, 1], 2], 3]$. The solution is not unique; also $x_{0, 1}^{0} = 1$, $x_{0, 2}^{1} = 1$, and $x_{2, 3}^{2} = 1$ produces the same plan with the same cost.
\end{example}

\subsection{Precise cost function as HUBO}\label{subsection:cost_function}

After defining the binary variables of our optimization model, we describe how we encode the cost function as a higher-order unconstrained binary optimization problem whose minimum is the optimal cost up to cross products for the left-deep join order selection problem. We describe the cost constraint first since the validity constraints can be computed based on terms that we compute for the cost constraint. First, we demonstrate the intuition behind the construction with an example.

\begin{example}\label{ex:example2}
Every join should be performed at exactly one rank for left-deep join trees. Starting from rank $0$, let us say that we choose to perform a join between the relations $R_1$ and $R_2$ and obtain $R_1 \bowtie R_2$. The corresponding activated binary variable is $x_{1,2}^{0} = 1$. Based on Def.~\eqref{eq:cost_function} of the cost function, the cost of performing this join is $f_{1,2}|R_1||R_2|$. Thus, if we decide to make this join at this rank, we include the term
\begin{equation*}
    f_{1,2}|R_1||R_2|x_{1,2}^{0}
\end{equation*}
to the cost HUBO. This example demonstrates that it is easy to encode the costs at rank $0$, which correspond to linear variables in the cost HUBO.

Next, we assume the query graph gives us a join predicate with selectivity $f_{2,3}$ between the tables $R_2$ and $R_3$. Now we ask how expensive it is to perform the join between intermediate result $R_1 \bowtie R_2$ and relation $R_3$. By Def.~\eqref{eq:cost_function}, the cost of making this join is
\begin{equation}\label{eq:card_123}
    f_{1,2}f_{2,3}|R_1||R_2||R_3|
\end{equation}
assuming that there is no edge $(R_1, R_3)$ which indicates that $f_{1,3} = 1$. Note that this is not the total cost of performing all the joins but the cardinality of the resulting table $R_1 \bowtie R_2 \bowtie R_3$. The left-deep join tree $(R_1 \bowtie R_2) \bowtie R_3$ should be selected if the following total cost function evaluates to a relatively small value
\begin{equation*}
    f_{1,2}|R_1||R_2|x_{1,2}^{0} + f_{1,2}f_{2,3}|R_1||R_2||R_3|x_{1,2}^{0}x_{2,3}^{1}.
\end{equation*}
When the binary variables are active, i.e., $x_{1,2}^{0} = x_{2,3}^{1} = 1$, the previous function evaluates the total cost of performing the join $(R_1 \bowtie R_2) \bowtie R_3$.

One of the key ideas is that the cardinality in Eq. \eqref{eq:card_123} does not depend on the join order but only on the tables that are part of the join at that point. In other words, this means that the cardinality in Eq. \eqref{eq:card_123} is the same for any join result that includes the relations $R_1$, $R_2$ and $R_3$ such as $(R_1 \bowtie R_3) \bowtie R_2$ and $R_1 \bowtie (R_2 \bowtie R_3)$. This naturally generalizes to any number of relations. The total costs of these plans likely differ because intermediate steps have different costs. Intuitively, our HUBO model seeks the optimal configuration of joins to construct the full join tree so that the sum of the intermediate results is the smallest.
\end{example}

Next, we formally describe constructing the HUBO problem that encodes the precise cost function for a complete left-deep join order selection problem respecting the structure of a given query graph $G$. The HUBO problem is constructed recursively with respect to the rank $r$. The construction of the HUBO problem becomes recursive because the definition of the cost function \eqref{eq:cost_function} is recursive.

\textbf{Step $r = 0$.} Let $G = (V, E)$ be a query graph. Based on Def.~\eqref{eq:cost_function}, we include the costs of making the rank $0$ joins to the cost HUBO. Precisely, we add terms
\begin{equation*}
    |R_i \bowtie R_j|x_{i,j}^{0} = f_{ij}|R_i||R_j|x_{i,j}^{0} = \alpha_{(i,j)}x_{i,j}^{0},
\end{equation*}
for every join $(R_i, R_j) \in E$, where we denote $\alpha_{(i,j)} := f_{ij}|R_i||R_j|$ the coefficient.

\textbf{Step $r = 1$.} For clarity, we also show step $r = 1$. Assuming we have completed step $r = 0$, we consider adding variables of type $x_{i,j}^{1}$. For every join $(R_i, R_j) \in E$, we select the adjacent joins $(R_{i'}, R_{j'})$ in the query graph. An adjacent join means that the joins share exactly one common table. This creates quadratic terms of type $x_{i,j}^{0}x_{i',j'}^{1}$ with coefficients of type $$\alpha^{1}_{(i,j,i',j')} = f_{ij}f_{i'j'}f_{i'j}f_{ij'}|R_i||R_j||R_i'||R_j'|.$$ So, we add terms $\alpha_{(i,j,i',j')}x_{i,j}^{0}x_{i',j'}^{1}$ to the cost HUBO.

\textbf{Step for arbitrary $r$.} Next, we consider adding a general rank $0 < r < |V| - 1$ variables of form $x_{i', j'}^{r}$ to the HUBO problem. For the general case, we formalize the method using connected subgraphs in the query graph: Let $\mathcal{S}$ be the set of size $r-1$ connected subgraphs in the query graph so that every subgraph corresponds to terms that were generated at step $r-1$. Since the cost function is defined recursively, this is an alternative way to express that we have added the variables to rank $r - 1$. This means that the HUBO problem, encoding the total cost up to this step, has the form:
\begin{align*}
    \underbrace{\sum_{(i, j) \in E}\alpha_{(i,j)} x_{i,j}^{0}}_{\text{case } r = 0} + \underbrace{\sum_{(i, j) \in E}\sum_{(i', j') \in E} \alpha_{(i,j,i',j')} x_{i,j}^{0}x_{i', j'}^{1}}_{\text{case } r = 1} + \ldots  
    + \underbrace{\sum_{S \in \mathcal{S}}\alpha_{S}\prod_{(R_i, R_j) \in S, 0\leq k \leq r - 1} x_{i,j}^{k}}_{\text{case } r - 1}.
\end{align*}
For simplicity, we first focus on generating the next term without a coefficient. Let $S \in \mathcal{S}$ be a fixed subgraph of the query graph. Let $(R_{i'}, R_{j'}) \in E$ be an edge that is not part of the subgraph $S$ but connected to it so that either $R_{i'} \in S$ or $R_{j'} \in S$ (but not both $R_{i'}, R_{j'} \in S$). This means we join exactly one new table. For this fixed subgraph $S$ and fixed join $(R_{i'}, R_{j'})$, we are going to add the following element to the cost HUBO:
\begin{equation}\label{eq:new_term}
\underbrace{\prod_{(R_i, R_j) \in S, 0\leq k \leq r - 1} x_{i,j}^{k}}_{\text{term at rank } r - 1} \underbrace{x_{i',j'}^{r}}_{\text{new variable at rank } r}
\end{equation}
The new term is just the ''old'' term from the previous step multiplied by the new variable $x_{i',j'}^{r}$.

\begin{figure}
\centering
\begin{subfigure}[t]{.49\linewidth}
  \centering
  \includegraphics[width = \textwidth]{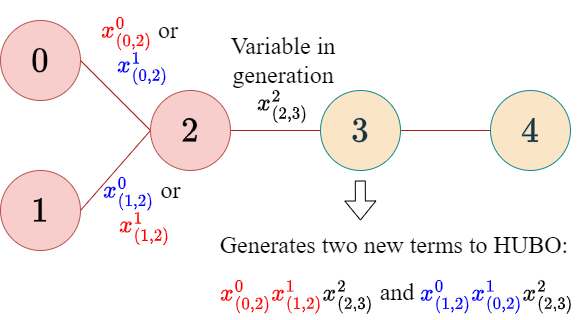}
  \caption{With relations 0, 1, and 2 already joined, relation 3 is the only option for left-deep plans. Two combinations arise based on whether edge $(0,2)$ or $(1,2)$ joins first. Coloring links graph variables, generating terms in HUBO.}
  \label{fig:subgraph_generation}
  \Description[subgraph_generation]{Assuming we have considered the relations 0, 1, and 2 at earlier ranks, then the only possible relation we can join in left-deep plans is relation 3. Two combinations are included depending on whether edge $(0,2)$ or edge $(1,2)$ is joined first. Coloring highlights the connection between variables in the graph and produces terms in HUBO.}
\end{subfigure}\hfill
\begin{subfigure}[t]{.49\linewidth}
  \centering
    \includegraphics[width = \textwidth]{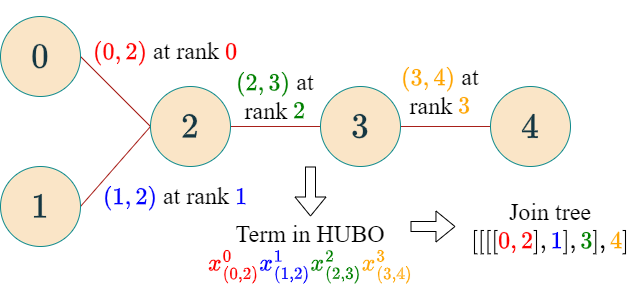}
    \caption{The figure presents one option as the final assignment for the binary variables in the query graph. The algorithm returns the corresponding join tree if the variables are selected true.}
    \label{fig:subgraph_to_terms}
    \Description[subgraph_to_terms]{The figure presents one option as the final assignment for the binary variables in the query graph. The algorithm returns the corresponding join tree if the variables are selected true.}
\end{subfigure}
\caption{Two examples for variable generation in tree query graph}
\label{fig:test}
\end{figure}

The new coefficient is easy to compute based on the subgraph $S$ and the latest included join $(R_{i'}, R_{j'})$. Precisely, consider the new induced subgraph $S' = S \cup \left\{(R_{i'}, R_{j'})\right\}$. Note that the induced subgraph $S'$ may contain new edges besides the latest included edge $(R_{i'}, R_{j'})$. However, the new coefficient for term \eqref{eq:new_term} is
\begin{equation}\label{eq:term_coefficient}
\alpha_{S'} = \prod_{(R_i, R_j) \in S'}f_{i,j}\prod_{R_i \in S'}|R_i|.
\end{equation}
This formula is just the general expression of Eq.~\eqref{eq:card_123} in Example \ref{ex:example2}. The previous construction is repeated for each subgraph $S \in \mathcal{S}$ and each edge adjacent to subgraph $S$ sharing one common vertex (i.e., a table). The idea of how the terms are appended at rank $2$ is visualized in Fig.~\ref{fig:subgraph_generation}. One of the final configurations is visualized in Fig.~\ref{fig:subgraph_to_terms}, which also shows how the terms are interpreted as join trees.

The algorithm to generate the cost HUBO is presented in Alg.~\ref{alg:hubo_term_construction}. It inputs a query graph and outputs a dictionary called HUBO, which stores the terms of the cost HUBO problem. The keys in this dictionary are sets of tables, and the values are sets of terms. We exclude the coefficient computation for simplicity since the coefficients can be efficiently computed with Eq.~\eqref{eq:term_coefficient}.

\begin{algorithm}
\caption{Construct Terms for Precise Cost Function HUBO}
\label{alg:hubo_term_construction}
\begin{algorithmic}[1]
\Require query\_graph
\Ensure terms for cost HUBO
\For{\textbf{each} $r$ \textbf{in} $0, \ldots, |V| - 2$}
    \For{edge \textbf{in} query\_graph}
        \State $R_i$, $R_j \gets$ edge[0], edge[1]
        \If{$r$ $=$ 0}
            \State HUBO[$\{R_i, R_j\}$] $\gets$ $x_{ij}^{0}$
        \Else
            \State joined\_tables $\gets$ [tables if $r - 1$ var in HUBO[tables]]
            \For{joined \textbf{in} joined\_tables}
                \State cond1 = $R_i \in$ joined and $R_j \notin$ joined
                \State cond2 = $R_j \in$ joined and $R_i \notin$ joined
                \If{cond1 \textbf{or} cond2}
                    \State new\_joined $\gets$ joined $\cup$ $\{R_i, R_j\}$
                    \State old\_terms $\gets$ HUBO[joined]
                    \State new\_terms $\gets$ multiply old\_terms with $x_{ij}^{r}$
                    \State HUBO[new\_joined] $\gets$ new\_terms
                \EndIf
            \EndFor
        \EndIf
    \EndFor
\EndFor
\end{algorithmic}
\end{algorithm}

\subsection{Encoding heuristic cost function as HUBO}\label{subsection:heuristic_cost_function}

While constructing the cost objective that encodes the optimal solution for the left-deep join order selection problem adds valuable precision, it can require computational resources as complexity increases. Thus, we have modified the cost function generation. We have included a greedy heuristic in the HUBO construction process (Alg.~\ref{alg:hubo_term_construction}) to include only those higher-order terms that are likely to introduce the most negligible cost to the total cost function.

The idea behind the heuristic is the following. First, we again include all the rank $0$ terms. When we start including rank $1$ terms, we consider only those rank $0$ terms whose cardinality (i.e., the coefficient in the HUBO objective) is minimal. We have included a tunable hyperparameter $n$ that selects $n$ terms with the smallest coefficients. Then, the HUBO construction continues with the selected subset of terms. With this heuristic, the size of the optimization problem is reduced remarkably, although we lose the guarantee of finding the optimal plan. 

The algorithm for this generation is almost identical to Alg.~\ref{alg:hubo_term_construction} except that line 7 should be changed from the current version to include only those table configurations with the minimum coefficient in the cost HUBO. In other words, we change line 7 to be ''$n$ many tables associated with rank $r - 1$ variable with the smallest coefficients''. We will later prove that this formulation produces at least as good a plan as the classical greedy algorithm and likely produces better for larger values $n$.
\section{Join order validity as HUBO}

HUBO problems, like QUBO problems, are required to return valid solutions. In this case, validity means that the join tree adheres to the query graph. All valid solutions are usable, although they might not minimize the cost. In this section, we present two approaches to encoding the validity of solutions: cost function dependent and independent. A cost-function-dependent approach is easy to construct but produces a larger number of higher-order terms. The cost-function-independent approach is closer to the standard QUBO formulations and identifies a collection of constraints the formulation needs to satisfy. The advantage of the second formulation is that the terms are primarily quadratic, which is easier to optimize in practice.

\subsection{Cost-function dependent validity}\label{subsubsection:cost_function_dependent}

Considering the cost function generation in the previous subsections, we have generated higher-order terms that encode valid join trees at rank $r = |V| - 2$. Considering the Alg.~\ref{alg:hubo_term_construction}, we can access these terms with $\texttt{HUBO[set of all relations]}$. Let us denote this set of terms as $H$. For example, one of those terms is represented in Fig.~\ref{fig:subgraph_generation}. The first validity constraint forces the model to select exactly one of these terms as true, which requires the final join tree to contain all the relations.

\textbf{Select one valid plan.} We utilize a generalized one-hot (or $k$-hot) encoding from QUBO formulations \cite{lucas_2014, Schonberger_Trummer_Mauerer_2023}. The encoding constructs a constraint that reaches its minimum when precisely $k$ variables are selected to be true from a given set of binary variables. In the generalized formulation, we construct an objective minimized when exactly $k$ terms are selected true from a set of higher-order terms. Let $H$ be the set of higher-order terms at rank $r = |V| - 2$. This functionality generalizes to higher-order cases with the following formulation:
\begin{equation}\label{eq:hubo_combinations}
    H_0 = \left( 1 - \sum_{h \in H} h \right)^2,
\end{equation}
where $h = \prod_{r = 0}^{|V| - 2} x_{i_r, j_r}^{r}$ for some indices $i_r$ and $j_r$ for $r = 0, \ldots, |V| - 2$ that depend on the query graph. The objective $H_0$ is always non-negative since it is squared. It is positive except when exactly one of the terms in the sum $\sum_{h \in H} h$ is $1$ when $H_0$ evaluates to $0$. The sum $\sum_{h \in H} h$ evaluates to $1$ when there is exactly one term $h$ such that all the variables in the product $\prod_{r = 0}^{|V| - 2} x_{i_r, j_r}^{r}$ are true. This combination is the valid join tree that the objective function returns as a solution to the minimization problem. Since these terms $h \in H$ have already been generated for the cost HUBO, we create this constraint based on them. Thus, we call this validity constraint cost-dependent.

\textbf{Every rank must appear exactly once in the solution.}
It is still possible to obtain solutions that include unnecessary true variables that do not affect the minimum of the final HUBO function. For example, consider the plan in Fig.~\ref{fig:subgraph_generation}. The HUBO that encodes this plan would have the same minimum even if we set that $x_{0,2}^{1} = 1$ because this variable would always be multiplied with variables set to $0$, and thus, activating a single variable would not affect the total cost. While we can solve this problem with classical post-processing, we still decided to fix it in the model itself. We include a constraint that every rank should appear exactly once in the solution. This is also an instance of one-hot encoding \cite{dimod_generators_combinations} and encoded with the following quadratic objective:
\begin{equation}\label{eq:every_rank_has_one_join}
    H_1 = \sum_{r = 0}^{|V| - 2}\left( 1 - \sum_{(R_i, R_j) \in E} x_{i,j}^{r} \right)^2.
\end{equation}
The objective $H_1$ is minimized when exactly one variable of type $x_{i,j}^{r}$ is selected to be true for each $0 \leq r <|V|- 1$.
\subsection{Cost-function independent validity}\label{subsubsection:cost_function_independent}

Although validity constraint \eqref{eq:hubo_combinations} is theoretically correct, it produces higher-degree terms due to exponentiation to the power of two, which we might want to avoid. Hence, we develop join tree validity constraints independently from the cost function. Notably, these validity constraints are often quadratic, i.e., QUBOs, and automatically supported by many solvers. Because we develop the theory considering the query graph's structure, we have slightly different constraints depending on the query graphs.

\subsubsection*{Clique graphs}
\textbf{Every rank must appear exactly once in the solution.} The first constraint is what we presented in Eq.~\eqref{eq:every_rank_has_one_join}. This constraint encodes that we perform exactly one join at every rank.

\textbf{Select connected, left-deep join tree.} The second constraint encodes that we penalize cases that do not form a connected, left-deep join tree. While it is clear that join trees must be connected, we must also encode that they are left deep because the cost HUBO does not evaluate bushy trees correctly. To achieve this, we include a constraint of the form
\begin{equation}\label{eq:respect_clique_graph}
    \sum_{r = 0}^{|V| - 2}\sum_{(i,j) \in E}\sum_{(i',j') \in E}Cx_{i,j}^{r}x_{i',j'}^{r+1},
\end{equation}
where the terms are included if $i \neq i'$ and $j \neq j'$, or $i = i'$ and $j = j'$, which means that we penalize these cases by increasing the objective's value by $C$. This makes the model favor cases where consecutive joins share one ''old'' table and include exactly one ''new'' table in the result. The penalizing term $C$ should be set high enough, and we will define its value later.

\textbf{Result contains all the tables.} The third constraint forces the fact that we join all the tables. Our framework identifies joins as pairs of tables $(R_i, R_j)$. This leads to one table $R_i$ appearing multiple times in variables, referring to different joins in a valid solution. An example of this is presented in Fig.~\ref{fig:subgraph_to_terms}, where table $2$ appears multiple times in the solution consisting of joins such as $(0,2)$ and $(1,2)$. This third constraint encodes that we count the number of tables and require that the count is at least one for each table. Counting tables requires minimally a logarithmic number of slack variables (the method to do the logarithmic encoding is presented in \cite{lucas_2014}) in terms of tables. For technical simplicity, we present the less efficient but equivalent method here:
\begin{equation*}
    \sum_{R_i}\left( 1 + \sum_{k = 2}^{|V| - 2}ky^{i}_{k} - \sum_{i}\sum_{r = 1}^{|V| - 2} x_{i,j}^{r} \right)^2.
\end{equation*}
The constraint reaches 0 if at least one variable of type $x_{i,j}^{r}$ is true for each table $R_i$. The other accepted cases are that we activate any number of $k \in \left\{ 2, \ldots, |V| - 2 \right\}$ many variables of type $x_{i,j}^{r}$ using the slack variables $y^{i}_{k}$. This means we must select at least one table but possibly $k$ many tables.
\subsubsection*{Chain, star, cycle, and tree graphs}

\textbf{At every rank, we have performed rank + 1 many joins.}
This constraint is related to the first validity constraint presented in \cite{Schonberger_Scherzinger_Mauerer,10.14778/3632093.3632112}. They present the constraint in terms of tables, whereas we have constructed it in terms of joins. The constraint encodes how many joins we must perform cumulatively at each rank. In other words, when rank is $0$, we select one variable of type $x_{i,j}^{0}$ to be true. When rank is $1$, we choose two variables of type $x_{i,j}^{1}$ to be true. Formally, this constraint is
\begin{equation}\label{eq:at_every_rank_select_rank_many_joins}
    \sum_{r = 0}^{|V| - 2}\left( r + 1 - \sum_{(i,j) \in E} x_{i,j}^{r} \right)^2.
\end{equation}
The constraint is minimized at $0$ when for each $0 \leq r < |V| - 1$, we have selected exactly $r + 1$ variables of type $x_{i,j}^{r}$ to be true.

\textbf{Include the previous joins in the proceeding ranks.}
This constraint is again similar to the second constraint presented in \cite{Schonberger_Scherzinger_Mauerer,10.14778/3632093.3632112} except that we express the constraint using joins. If the join happened at rank $r$, it should be included in every proceeding rank $\geq r$. In other words, we keep the information of the performed joins to the following ranks. This can be achieved with the following constraint
\begin{equation}\label{eq:select_same_join_for_proceeding_ranks}
    \sum_{(i,j) \in E}\sum_{r = 1}^{|V| - 2} x_{i,j}^{r - 1}(1 - x_{i,j}^{r}).
\end{equation}
Now, if $x_{i,j}^{r - 1}$ is active, then the model favors the case that $x_{i,j}^{r}$ is active too since in that case $1 - x_{i,j}^{r} = 0$. If $x_{i,j}^{r} = 0$, the term evaluates to $1$, which is considered penalizing.

\textbf{Respect query graph: chain, star, cycle.}
Since the cost functions are designed to respect structures of query graphs, we use the following constraint to encode the graph structure in chain, star, and cycle graphs:
\begin{equation}\label{eq:respect_query_graph1}
    \sum_{r = 0}^{|V| - 1}\sum_{(i,j) \in E}\sum_{(i',j') \in E} -C x_{i,j}^{r}x_{i',j'}^{r},
\end{equation}
where $|\left\{ i, j \right\} \cap \left\{ i', j' \right\}| = 1$, which means that the joins have to share exactly one table. Setting the coefficient $-C$ as negative, we favor the cases when the joins share precisely one table. This constraint is complementary to constraint \eqref{eq:respect_clique_graph}.

\textbf{Respect query graph: tree.}
Unfortunately, the previous constraint \eqref{eq:respect_query_graph1} fails to encode the minimum for certain proper trees, which contain nodes with at least three different degrees. The simplest, problematic tree shape is represented in Figure \ref{fig:subgraph_to_terms} (node 2 has degree 3, node 3 has degree 2, and the others have degree 1). The problem is that with constraint \eqref{eq:respect_query_graph1}, not all the join trees have the same minimum energy due to nodes' different degrees. To address this problem, we develop an alternative constraint
\begin{equation*}
    \sum_{r = 1}^{|V| - 1} \left( r - \sum_{(i,j) \in E}\sum_{(i',j') \in E} x_{i,j}^{r}x_{i',j'}^{r}\right)^2,
\end{equation*}
where again we form the sum over the elements if the indices satisfy $|\left\{ i, j \right\} \cap \left\{ i', j' \right\}| = 1$. At every rank, this constraint selects $r$-many pairs of type $x_{i,j}^{r}x_{i',j'}^{r}$ to be true. This forces the returned join tree to respect the query graph and be connected. This constraint is slightly more complex than the previous constraints since it is not a quadratic but a higher-order constraint that includes terms of four variables. On the other hand, it does not introduce additional variables.

\textbf{Scaling cost and validity.} Finally, we have to scale cost HUBO $H_{\text{cost}}$ and validity objective $H_{\text{val}}$ properly so that we favor valid solutions over minimizing cost:
\begin{displaymath}
H_{\text{full}} = H_{\text{cost}} + CH_{\text{val}}.
\end{displaymath}
We noticed that a value that worked consistently in practice is $C = H_{\text{cost}}(x)$ where $x = (1, \ldots, 1)$ is a binary vector containing only ones and $H_{\text{cost}}$ is normalized so that the coefficients are in the interval $(0,1]$.
\section{Theoretical analysis}
We prove two theorems that give bounds for the quality of the solutions, which are expected to be reached with the cost functions. We have the same initial assumptions for both proofs, which we state next. We consider that dynamic programming (DP) and greedy algorithms perform as many steps as we perform joins. We assume the reader is familiar with these two algorithms but have included their definitions in the appendix. 

By Def.~\eqref{def:binary_variables} of the binary variables, the number of steps coincides with the rank index in the cost functions. Thus, we can prove the theorems by induction on the rank parameter $r$ in the proposed algorithms, as well as the steps in the DP and greedy algorithms, by showing that for each $r$, there exists a solution from HUBO that corresponds to the same plan computed by the classical algorithms.

\begin{theorem}\label{thm:dp_bound}
Let  $H_{\text{cost}}$ be the cost HUBO defined in Subsection \ref{subsection:cost_function} and let $H_{\text{val}}$ be the binary formulation for the join order validity constraints. Let $x$ be the point that minimizes the full problem $H_{\text{cost}} + CH_{\text{val}}$. Then, the join order cost $H_{\text{cost}}(x)$ is equal to the cost computed by the dynamic programming algorithm without cross-products.
\end{theorem}
\begin{proof}
First, assume that the rank is $0$, and we are at the first iteration in the dynamic programming (DP) algorithm. By Alg.~\ref{alg:hubo_term_construction} (line 5), we include all the joins between the leaf tables to the cost function $H_{\text{cost}}$. Considering the DP algorithm, we compute the same costs for joining the leaf tables and include the combinations and costs in the DP table. Thus, the HUBO encodes the same plans as the DP algorithm.

Let us assume that we are at rank $r > 0$, and we have applied Algorithm \ref{alg:hubo_term_construction} and the DP algorithm up to $r - 1$ steps. The DP algorithm considers all the intermediate joins from the previous step (i.e., at rank $r-1$). For each of these joins, it performs the possible join for each table that is not yet included in the intermediate result with respect to the query graph. These results are kept in the DP table for the next iteration. While the DP algorithm keeps the total cost of each intermediate result in the DP table, our cost objective $H_{\text{cost}}$ encodes only the ''local'' intermediate costs, which are computed with Eq.~\eqref{eq:term_coefficient}. Based on Algorithm \ref{alg:hubo_term_construction}, we take the terms of rank $r-1$ and compute the corresponding new terms (i.e., intermediate join plans) and coefficients (i.e., intermediate join costs) for each table that is not yet included in the intermediate result with respect to the query graph. Instead of computing the total cost, we add the terms and coefficients to the cost objective $H_{\text{cost}}$. This process encodes the same cost that is stored in the DP table because we can choose to activate those $H_{\text{cost}}$ terms, which produce the total cost for each value stored in the DP table. This leads to the claim that there is a point where $H_{\text{cost}}$ achieves the same cost as the DP algorithm.
\end{proof}

\begin{theorem}\label{thm:greedy_bound}
Let $H_{\text{cost}}$ be the heuristic method's cost HUBO defined in Subsection \ref{subsection:heuristic_cost_function}, and let $H_{\text{val}}$ be the binary formulation for the join order validity constraints. Let $x$ be the point that minimizes the full problem $H_{\text{cost}} + CH_{\text{val}}$. Let $C_{\text{greedy}}$ be the cost computed by the greedy algorithm without cross-products. Then, $H(x) \leq C_{\text{greedy}}$, i.e., the cost from the greedy algorithm gives an upper bound for the cost from the heuristic algorithm.
\end{theorem}
\begin{proof}
First, assume that the rank is $0$, and we are at the first iteration in the greedy algorithm. By the definition of the heuristic method in Subsection \ref{subsection:heuristic_cost_function}, we include all terms corresponding to the joins between leaf tables to the cost function $H_{\text{cost}}$. The greedy algorithm includes only one join with the minimum cost to the join tree at the same step. Thus, our heuristic method encodes the first step of the greedy algorithm.

Let us consider a general rank $r > 0$ and assume that we have applied the heuristic HUBO construction and the greedy algorithm up to $r - 1$ steps. By the definition of our heuristic method, we select a subset of $n$-many table combinations from the previous rank $r - 1$ with the smallest coefficients. Simultaneously, in the greedy algorithm, we compute the costs of joining the previous step's intermediate result with the possible joins that are left with respect to the query graph and keep the join tree with the minimum total cost. This minimum total cost is always achieved by including the join tree corresponding to the terms with the smallest coefficient in the $H_{\text{cost}}$ at rank $r$. Since the terms with the smallest coefficients correspond to the cheapest plans, the heuristic algorithm will encode the same plan that the greedy algorithm finds. For larger numbers of $n$, $H_{\text{cost}}$ will also include other intermediate results. Since our heuristic method and the greedy algorithm keep the join plans with the minimum cost at each step, we can deduce that our algorithm encodes the same plan (and others depending on value $n$) as the greedy algorithm. This leads to the claim that $H(x) \leq C_{\text{greedy}}$.
\end{proof}

\begin{figure}
    \centering
    \includegraphics[width=0.4\linewidth]{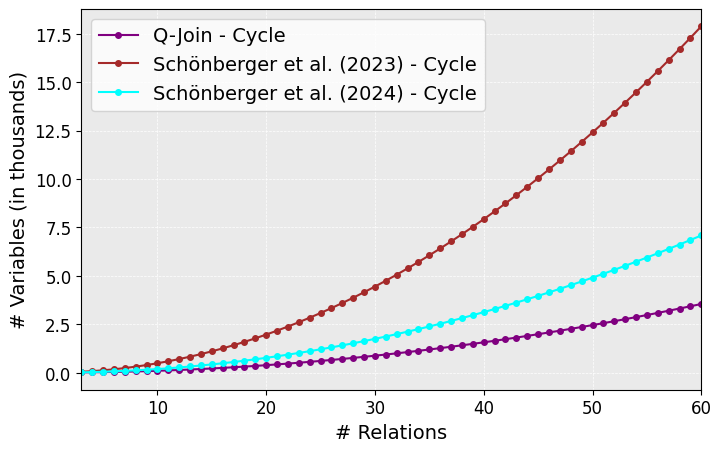}
    \caption{We compare the number of mandatory variables in \cite{Schonberger_Scherzinger_Mauerer,10.14778/3632093.3632112} to all variables in our optimization model.}
    \label{fig:variable_scalability_cycle}
    \Description[Comparison of variable scalability between this and previous methods]{The plot shows that our method has the best variable scalability compared to the previous methods.}
\end{figure}

Besides the theorems, our method archives advantageous variable scalability compared to the most scalable methods \cite{Schonberger_Scherzinger_Mauerer, 10.14778/3632093.3632112}. The model in \cite{DBLP:conf/q-data/SaxenaSS24} uses the same variable definitions as \cite{10.14778/3632093.3632112}, so the scalability comparison also applies to this paper as well. The scalability is visualized in Fig.~\ref{fig:variable_scalability_cycle} for cycle query graphs. The chain, star, and tree graphs have identical relative scalability in all three cases. We want to point out that these are the only mandatory variables for the two compared methods. The previous techniques require more variables to estimate the cost thresholds, which depend on the problem. We excluded the variable scalability of \cite{Nayak_Winker_Groppe_Groppe_2024} since the growth of variable count is exponential in their work.

\section{Discussion}

The starting point for our work has been primarily the previous quantum computing formulations \cite{Schonberger_Scherzinger_Mauerer, Winker_Calikyilmaz_Gruenwald_Groppe_2023, Schonberger_Trummer_Mauerer_2023, Nayak_Winker_Groppe_Groppe_2024, Franz_Winker_Groppe_Mauerer_2024,10.14778/3632093.3632112, DBLP:conf/q-data/SaxenaSS24} for join order selection problem. Compared to previous quantum-based join order selection research, our method has a formal guarantee of the quality of the results; our cost function models the intermediate results more accurately, and the number of binary variables is the lowest. We have also presented a fundamentally different approach to the well-researched problem since the previous methods often relied on the MILP formulation \cite{Trummer_Koch_2017}.

We have performed an extensive experimental evaluation, and the results are shown in the appendix. The experimental results demonstrate that the Theorems \ref{thm:dp_bound} and \ref{thm:greedy_bound} are respected well in practice. Our methods' scalability outperforms many previous works \cite{Schonberger_Scherzinger_Mauerer, Franz_Winker_Groppe_Mauerer_2024, Schonberger_Trummer_Mauerer_2023}, where the authors have demonstrated their algorithms with 2 to 7 relations. The method proposed in \cite{10.14778/3632093.3632112} has the best scalability and accuracy, but the algorithm lacks a guarantee of optimality and uses more variables. Still, the comparison to this approach shows accuracy similar to our method. Our evaluation also explores the method in \cite{10.14778/3632093.3632112} for the query graphs not studied in the original paper. The detailed results are in the appendix. We also note that classical solvers compete with the quantum annealers in this task even though they run locally on a laptop. This demonstrates that quantum annealers are not scalable enough to solve arbitrary problems, but their performance is crucially problem-dependent.

The biggest challenge in our method is the higher-order terms. Very few methods can effectively tackle HUBO optimization problems. In this regard, quantum computing appears to be theoretically one of the most promising approaches to optimizing complex HUBOs. In future research, we are excited to study universal quantum computing capabilities to solve HUBOs.

\section{Conclusion and future work}

In this work, we have developed three novel higher-order binary optimization models to solve join order selection problem. The HUBO problems can be divided into cost function and validity constraints. We presented two new binary optimization formulations for the cost function and proved that one encodes the same join trees as the dynamic programming algorithm without cross-products. The other find plans that are at least as good as those returned by the greedy algorithm without cross-products. Finally, we presented a comprehensive experimental evaluation of these algorithms on various quantum and classical solvers. The experimental evaluation demonstrated the practical scalability of this algorithm and the fact that we respect the proven bounds in practice.

Previous methods have been limited to inner joins. Extending our binary variables to model non-inner joins is theoretically straightforward, for instance, by adding a component to indicate the join type (inner or outer). While this modification is simple, the cost function and constraints also need adjustments. Since our cost HUBO is explicitly based on the recursive join order cost function, encoding predicate dependencies is also feasible. Exploring these extensions offers promising directions for future research in join order optimization with quantum computing.

\bibliographystyle{ACM-Reference-Format}
\bibliography{sample-base}

\appendix
\section{Appendix: Background on quantum computing}

\subsection{Quantum circuit model}

In this part, we define the quantum circuit model \cite{Nielsen_Chuang_2010}. Classical computers operate with bits having discrete values of $0$ or $1$. Quantum computing is based on quantum bits, called qubits, that are formally represented as vectors in a complex-valued Hilbert space. For a single qubit system, the basis vectors of this space are $|0\rangle:= [1, 0]^{\top}$ and $|1\rangle:= [0, 1]^{\top}$. Every state in a single-qubit quantum computer can be represented as a complex-valued linear combination of these basis states as $|\varphi\rangle = \alpha |0\rangle + \beta |1\rangle$, where $\alpha, \beta \in \mathds{C}$ so that $|\alpha|^2 + |\beta|^2 = 1$. The values $\alpha$ and $\beta$ are also called amplitudes, and their squared lengths $|\alpha|^2 $ and $|\beta|^2$ can be interpreted as a probability distribution. Since a qubit is an element of a Hilbert space equipped with the standard tensor product, we can construct larger systems by applying tensor product operation between the smaller systems. For example, a two-qubit system is described by four basis states: $|0\rangle\otimes|0\rangle = |00\rangle$, $|01\rangle$, $|10 \rangle$ and $|11\rangle$.

The quantum algorithm is implemented by applying quantum logic gates to the qubits. These gates must respect the condition that the resulting quantum system has $\sum_{i}|\alpha_i|^2 = 1$ over the amplitudes $\alpha_i$. The operations which satisfy this property are the unitary matrices. A matrix $U$ over the complex numbers is unitary if its inverse is its conjugate transpose. The conjugate transpose is obtained by conjugating the matrice's complex-valued elements and then transposing the matrix. A quantum circuit represents a quantum computational system where the circuit's wires represent qubits, and operations are represented as gates acting on wires. 

The system is measured after the gates have been applied to the qubits. Measurement operation is especially characteristic of quantum computing and does not have a similar role in classical computing. Informally, measuring corresponds to the return statement at the end of a classical function. The most common measurement operation is measuring on a computational basis. This means that after a measurement, we obtain a classical bit string whose length is the number of qubits. For example, if we measure a 2-qubit system, the possible measurement results are $00$, $01$, $10$, and $11$. The connection to the amplitudes is the following: if the system has a state $\alpha_{11}|11\rangle$, then the value $|\alpha_{11}|^2$ is the probability that we obtain the result $11$ when we measure. 

In quantum computing, we can perform various measurements \cite{Nielsen_Chuang_2010}. In this work, we measure an expectation value of an observable, which is a typical measurement operation in practical applications. This measurement estimates the Hamiltonian's energy value corresponding to the cost we aim to minimize in the combinatorial optimization problem. The expectation value can be interpreted as the weighted average of all possible measurement outcomes, where each outcome is weighted by its probability. However, it does not necessarily correspond to the most probable measurement outcome. Formally, suppose $H$ is an operator such as a Hamiltonian. In that case, we compute $\langle \varphi | H | \varphi \rangle = \sum_{j}\lambda_j |\langle\varphi | \lambda_j\rangle|^2$, where $\lambda_j$ and $|\lambda_j \rangle$ are the eigenvalue and eigenvectors of the operator $H$ and the state $| \varphi \rangle$ is the state where the quantum mechanical system is before the measurement. Now the sum is the weighted average of the eigenvalues which are precisely the measurement results when measuring the observable: the eigenvalues $\lambda_j$ are the measurement outcomes, and $|\langle\varphi | \lambda_j\rangle|^2$ corresponds to the probability, which can be viewed as an overlap between the eigenstate $|\lambda_j \rangle$ and the system's current state $| \varphi \rangle$.

\subsection{QAOA and VQE}

On the gate-based universal quantum computers, we can apply the Quantum Approximate Optimization Algorithm (QAOA) \cite{farhi2014quantum} or Variational Quantum Eigensolver (VQE) \cite{Peruzzo_2014} to find the ground state of the Hamiltonian. Both QAOA and VQE are variational algorithms, meaning we execute them in two phases: first, we execute the circuit with a fixed parameter configuration on a quantum computer, then we estimate the circuit's gradient on a classical computer and tune the parameters. These two phases are repeated until a sufficiently good parameter configuration is found. 

\begin{figure}
    \centering
    \input{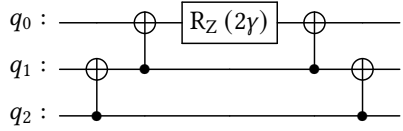}
    \caption{Circuit implementing the Hamiltonian $H = 2 \sigma_{z}^{0} \otimes \sigma_{z}^{1} \otimes \sigma_{z}^{2}$}
    \label{fig:circuit_example}
\end{figure}

First, we focus on QAOA. In this algorithm, we create two Hamiltonians: a mixer Hamiltonian $H_{\mathrm{mix}}$ and a problem Hamiltonian $H_{C}$. The most common mixer Hamiltonian is $H_{\mathrm{mix}} = \sum_{j}\sigma_{x}^{j}$. These Hamiltonians roughly correspond to the initial Hamiltonian and the problem Hamiltonian in Eq.~\eqref{eq:adiabatic_Hamiltonian}, and the cost Hamiltonian encodes the solution to the optimization problem. Then, we prepare two parametrized circuits for each Hamiltonian: $U_{C}:= e^{-i\gamma H_C}$ and $U_{\mathrm{mix}}:= e^{-i\alpha H_{\mathrm{mix}}}$. More concretely, implementing Hamiltonians as circuits, we follow the idea presented in \cite{Nielsen_Chuang_2010}. For example, if the Hamiltonian is $H = 2\sigma_{z}^{0} \otimes \sigma_{z}^{1} \otimes \sigma_{z}^{2}$, then the corresponding parametrized circuit is in Fig.~\ref{fig:circuit_example}. The important point is that we can natively encode and optimize HUBO problems using this method.
Next, we define the complete QAOA circuit, which is a repeated application of circuits $U_C$ and $U_{\mathrm{mix}}$ for $n$ times. The precise value of $n$ depends on the problem, but already values such as $3$ have proved to be good \cite{farhi2014quantum}. Then, we prepare an equal superposition over all the basis states, choose random initial values for the parameters $\gamma$ and $\alpha$, and apply the complete QAOA circuit to the state. The equal superposition encodes that initially, every possible binary configuration has an equal probability of being selected as a solution. After preparing the circuit, we measure the expectation value of the cost Hamiltonian $H_C$. Then, we repeat the circuit execution with modified parameters $\gamma$ and $\alpha$. Since parametrized unitaries are differentiable, the circuits are also differentiable with respect to the parameters, and we can optimize the parameters $\gamma$ and $\alpha$ using classical stochastic gradient descent methods. The optimization goal is to find a parameter configuration that minimizes the expectation value of the Hamiltonian, which solves the combinatorial optimization problem.

The basic principles of VQE are similar to those of QAOA, except that the VQE circuit does not implement $U_C$ and $U_{\mathrm{mix}}$ circuits. The parameterized circuit in VQE is a sophisticated guess that is optimized to minimize the energy of the Hamiltonian using classical gradient descent methods. In VQE, we also measure the expectation value of the cost Hamiltonian $H_C$. QAOA and VQE have many modified versions that tackle challenges in the current noisy intermediate-scale quantum computing. 
\section{Appendix: HUBO to QUBO reduction}

Next, we describe how to reduce HUBO problems to QUBO problems. This rewrite process is necessary to utilize a wider variety of optimization platforms. In this work, we mainly rely on the D-wave's Ocean framework's utility of automatically translating HUBO problems into QUBO problems. The translation is based on a polynomial reduction by minimum selection or by substitution \cite{polynomial_reductions}.

The HUBO to QUBO reduction based on minimum selection \cite{polynomial_reductions} follows the scheme
\begin{equation*}
    xyz = \max_{w}w(x+y+z-2),
\end{equation*}
which iteratively replaces the higher order terms $xyz$ with lower order terms by introducing auxiliary binary variables $w$. Depending on the order in which variable terms are replaced, the QUBO formulation may vary in format, and the number of binary variables, but the minimum point remains unchanged.

Rewriting mechanism by substitution utilizes the following formula
\begin{equation*}
    xyz = \min_{w}\left\{wz + \mathrm{MP}(x, y ; w)\right\},
\end{equation*}
where $M > 1$  is a penalty weight and $P$ is a penalty function that has the lowest value when $w = xy$. The details of why these rewriting methods reach the same minimum are explained in \cite{polynomial_reductions}. These reduction methods provide a technically easy method to encode HUBOs as QUBOs, but they also introduce auxiliary variables depending on the number of higher-order terms.
\section{Dynamic programming and greedy algorithms}

The dynamic programming and greedy algorithms implemented here are based on \cite{Neumann_course}. Dynamic programming for join order selection comprises a general class of approaches to optimize the join order selection. In our work, we have fixed the cost function (Eq.~\eqref{eq:cost_function}) and employed the dynamic programming algorithm with and without cross-products. The algorithm without cross-products is presented in Alg.~\ref{alg:dp-join-order}. It relies on functions that create left-deep trees for trees $T_1$ and $T_2$ and return costs for join trees based on the cost function in Eq.~\eqref{eq:cost_function}.

\begin{algorithm}
\caption{Dynamic Programming for Join Order Optimization With Cross-Products}
\label{alg:dp-join-order}
\begin{algorithmic}[1]
\Require relations $R = \{r_1, r_2, \dots, r_n\}$, selectivities $S$
\Ensure optimal left-deep join tree in $\text{dp\_table}[R]$
\State initialize $\text{dp\_table}$
\For{$r \in R$}
    \State $\text{dp\_table}[\{r\}] \gets r$ \Comment{Base case: single relation}
\EndFor
\For{$s = 2 \ \mathbf{ to } \ |R|$} \Comment{Size of subsets from 2 to $n$}
    \For{$\text{subset} \subseteq R$ such that $|\text{subset}| = s-1$}
        \If{$\text{subset} \in \text{dp\_table}$}
            \For{$r \in R \setminus \text{subset}$} \Comment{Relations not in subset}
                \State $T_1 \gets \text{dp\_table}[\text{subset}]$
                \State $T_2 \gets \text{dp\_table}[\{r\}]$
                \State $\text{T}, \text{T\_cost} \gets \text{create\_join\_tree}(T_1, T_2, R, S)$
                \State $\text{join\_key} \gets \text{subset} \cup \{r\}$
                \If{$\text{join\_key} \notin \text{dp\_table}$}
                    \State $\text{dp\_table}[\text{join\_key}] \gets T$ \Comment{Update if key not in table}
                \EndIf
                \If{$\text{join\_key} \in \text{dp\_table}$}
                    \State $\text{prev\_cost} \gets \textbf{cost}(\text{dp\_table}[\text{join\_key}], R, S)$
                    \If{$\text{T\_cost} < \text{prev\_cost}$}
                        \State $\text{dp\_table}[\text{join\_key}] \gets T$ \Comment{Update if lower cost}
                    \EndIf
                \EndIf
            \EndFor
        \EndIf
    \EndFor
\EndFor
\end{algorithmic}
\end{algorithm}

The algorithm that computes the dynamic programming result without cross-products is similar except that for a query graph $G$, we change line 6: \textbf{for} connected subgraph $\subset G$ such that $|\text{subgraph}| = s - 1$ \textbf{do}. Then, the algorithm proceeds with the connected subgraphs of size $s - 1$ instead of all subsets of size $s - 1$.

The greedy algorithm is the other standard algorithm to optimize join order selection, and we represent it in Alg.~\ref{alg:greedy-join-order}. Similarly to the dynamic programming algorithm, we can consider only solutions without cross-products so that we iterate only over tables connected to one of the tables already joined. In other words, at each step, we compute a value called adjacent\_tables which contains those tables $R_i$ so that if edge $(R_i, R_j) \in G$ in the query graph $G$, then we require that $R_i \notin \text{joined\_tables}$ but $R_j \in \text{joined\_tables}$. 

\begin{algorithm}
\caption{Greedy Algorithm for Join Order Selection}
\label{alg:greedy-join-order}
\begin{algorithmic}[1]
\Require relations $R = \left\{ r_1, \ldots, r_n \right\}$, selectivities $S$
\Ensure Greedy join order tree

\State $\text{join\_result} \gets [r_i, r_j]$ so that $f_{i,j}|r_j||r_j|$ is the smallest

\For{$1 \ \text{to} \ |R| - 1$}
    \State $\text{min\_cost} \gets \infty$
    \State $\text{min\_table} \gets \text{None}$
    \For{$\text{table} \in \text{relations} \setminus \text{joined\_result}$} \Comment{Iterate over tables which are not joined}
        \State $\text{current\_join\_tree} \gets [\text{table}, \text{join\_result}]$
        \State $\text{current\_cost} \gets \textbf{cost}(\text{current\_join\_tree}, R, S)$
        \If{$\text{current\_cost} < \text{min\_cost}$} \Comment{Choose table with smallest cost}
            \State $\text{min\_cost} \gets \text{current\_cost}$
            \State $\text{min\_table} \gets \text{table}$
        \EndIf
    \EndFor
    \State $\text{join\_result} \gets [\text{min\_table}, \text{join\_result}]$
\EndFor
\end{algorithmic}
\end{algorithm}
\section{Appendix: Experimental results}
In this section, we present the results of the comprehensive experimental evaluation, which contains various combinations of query graphs, problem formulations, and classical and quantum optimizers. For each method proposed in this work, we evaluated the technique against five common query graph types: clique, star, chain, cycle, and tree. Each query graph is labeled using the format 'Graph name - number of nodes'. For each graph type and graph size, we randomly selected 20 query graph instances with cardinalities and selectivities. The cardinalities are randomly sampled from the range 10 to 50 and selectivity from interval $(0, 1]$. The costs are summed over 20 runs, describing a realistic cumulative error, and scaled with respect to the cost returned from the dynamic programming algorithm \textit{with} cross-products, which is the optimal left-deep plan. The anonymized code for this experimental evaluation can be found on GitHub \cite{anonymous2024qjoin}. Since we have used 20 query graph instances for five different graph types of sizes 3 to 60 and solved them with four different quantum and classical solvers, the total number of evaluated query graphs is in the thousands.


We have decided to focus on the quality of solutions instead of optimization time. Although time is crucial in real-life cases, integrating quantum computational systems with classical systems still brings an unavoidable overhead. Quantum computers work at the time scale of nano and milliseconds, but the classical pre-and post-processing makes the total computation time relatively long in practice. Concretely, these pre-and post-processing phases are demonstrated by such steps as encoding problems in HUBO/QUBO format, submitting them to a quantum computer over a possibly slow connection, and even waiting in line for the quantum computer to become available from the other users. After executing the workload, we need to translate the obtained results back into a format that allows us to interpret them in the light of the original problem.

\textbf{Summary of proposed methods.} We have proposed three algorithms to solve the join order selection problem with a higher-order unconstrained binary optimization model. Table \ref{table:methods} introduces names for these methods, which are used in this section.

\begin{table}[!ht]
\centering
\resizebox{0.5\columnwidth}{!}{
\begin{tabular}{|c|c|c|c|}
\hline
\textbf{Method name} & \textbf{Cost function} & \textbf{Validity constraint} \\ \hline
precise 1 & precise cost function & cost function dependent \\ \hline
precise 2 & precise cost function & cost function independent \\ \hline
heuristic & heuristic cost function & cost function dependent \\ \hline
\end{tabular}
}
\caption{Summary of proposed algorithms}
\label{table:methods}
\end{table}

\subsection{Evaluating \textrm{Precise 1} formulation}

\begin{figure*}[tbh]
    \centering
    \includegraphics[width=\textwidth]{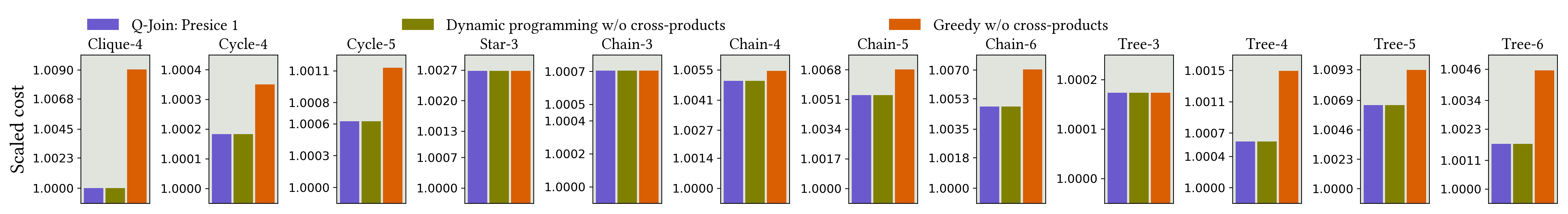}
    \caption{Precise 1 results using the D-Wave's exact poly solver}
    \label{fig:poly_solver_precise_1}
    \Description[Other results]{}
\end{figure*}

First, we evaluate \textrm{Precise 1} formulation, which combines precise cost function and cost-dependent validity constraints. Fig.~\ref{fig:poly_solver_precise_1} shows the results of optimizing join order selection using the D-Wave's exact poly solver. Following the bounds given by Theorem \ref{thm:dp_bound}, the HUBO model consistently generates a plan that matches the quality of the plan produced by the dynamic programming algorithm without the cross products. We also see that the returned plans are only at most $0.7\%$ bigger than the optimal plan from dynamic programming with the cross products. We have excluded some results where the HUBO model produced the exact optimal plan: Clique-3, Cycle-3, Star-4, and Star-5.

\begin{figure*}[tbh]
    \centering
    \includegraphics[width=\textwidth]{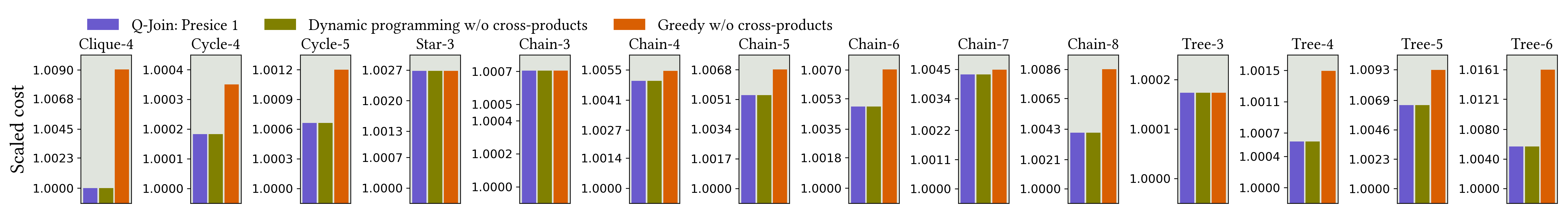}
    \caption{Precise 1 results using Gurobi solver}
    \label{fig:gurobi_precise_1}
    \Description[Other results]{}
\end{figure*}

Second, we solved the same HUBO formulations using a classical Gurobi solver after the HUBO problem was translated into the equivalent QUBO problem. The results are presented in Fig.~\ref{fig:gurobi_precise_1}. The HUBO to QUBO translation does not decrease the algorithm's quality, and Gurobi finds the correct plans. The results stay very close to the optimal join tree, always being as good as a dynamic programming algorithm without cross products.

Third, we solved the same problems using D-wave's Leap Hybrid solver, a quantum-classical optimization platform in the cloud. In this case, the results are consistently as good as those from the dynamic program algorithm without the cross products, with some exceptions due to the heuristic nature of the quantum computer: Cycle-6, Chain-7, and Tree-6.

\begin{figure*}[tbh]
    \centering
    \includegraphics[width=\textwidth]{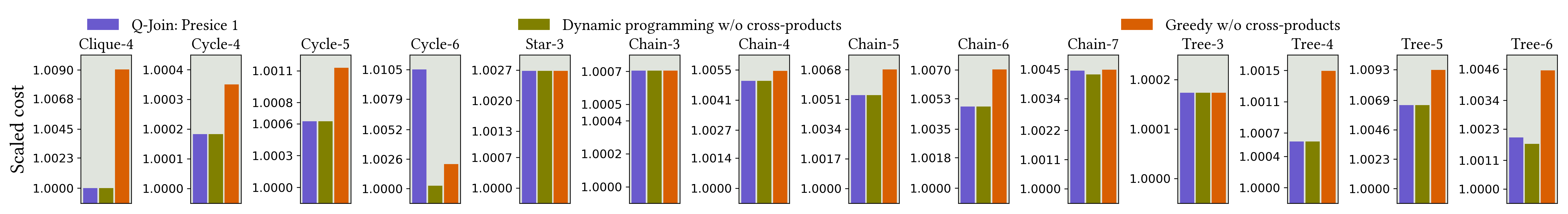}
    \caption{Precise 1 results using D-Wave's Leap Hybrid solver}
    \label{fig:leap_precise_1}
    \Description[Other results]{}
\end{figure*}

\begin{figure*}[tbh]
    \centering
    \includegraphics[width=\textwidth]{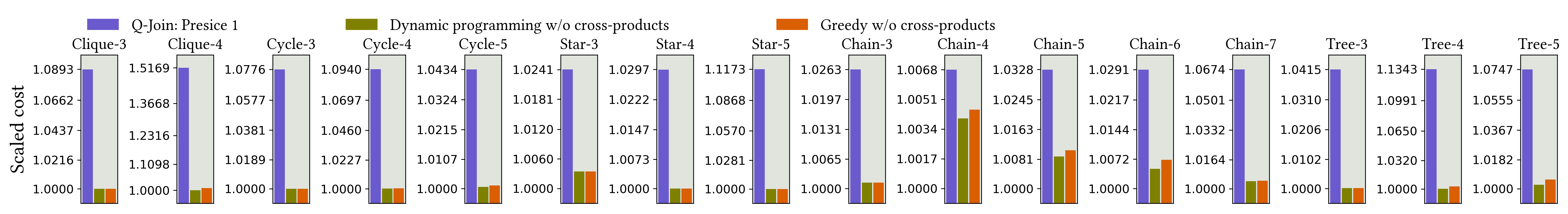}
    \caption{Precise 1 results using D-Wave's standard solver}
    \label{fig:dwave_precise_1}
    \Description[Other results]{Everything is included in these results}
\end{figure*}

Finally, Fig.~\ref{fig:dwave_precise_1} shows the results from D-Wave's quantum annealer, which does not utilize hybrid features to increase solution quality. This resulted in performance that did not match the performance of the previous solvers, and this performance decrease was already identified in \cite{Schonberger_Scherzinger_Mauerer}. While quality was not as good as the previous solutions, the valid plans were still usable, mainly only a few percent from the global optimal. 

\subsection{Evaluating \textrm{Precise 2} formulation}

The key idea behind the Precise 2 formulation is to tackle larger join order optimization cases because the validity constraints are more efficient regarding the number of higher-order terms. We include the exact poly solver results to demonstrate that this formulation encodes precisely the correct plans. For the other solvers, we only show results that optimized larger queries compared to the previous Presice 1 method.

\begin{figure*}[tbh]
    \centering
    \includegraphics[width=\textwidth]{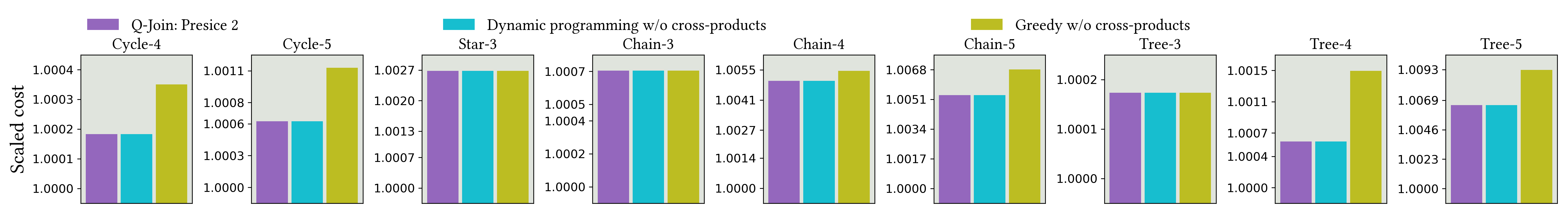}
    \caption{Precise 2 results using the D-Wave's exact poly solver}
    \label{fig:precise_2_exact_poly_solver}
    \Description[Precise 2 results using the D-Wave's exact poly solver]{}
\end{figure*}

First, the results from the exact poly solver in Fig.~\ref{fig:precise_2_exact_poly_solver} demonstrate that this algorithm follows the bounds of Theorem \ref{thm:dp_bound}. In practice, the returned plans are again very close to the optimal plans. We can also see that compared to the Precise 1 method, the different sets of validity constraints work equally well.

\begin{figure*}[tbh]
    \centering
    \includegraphics[width=\textwidth]{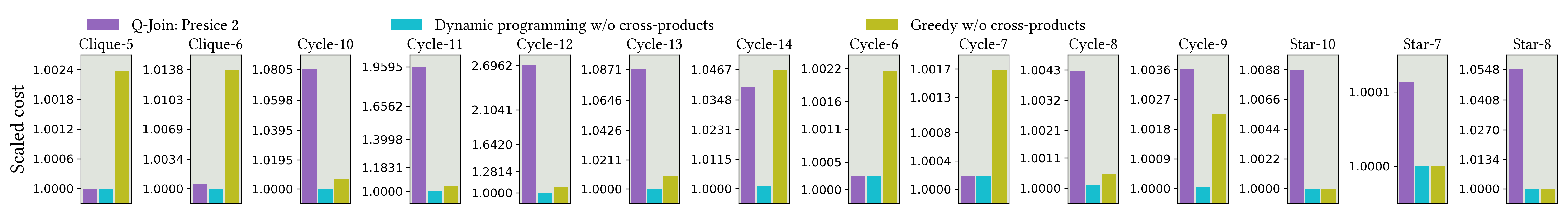}
    \caption{Precise 2 results using Gurobi solver}
    \label{fig:precise_2_gurobi_1}
    \Description[Precise 2 results using Gurobi solver]{}
\end{figure*}

\begin{figure*}[tbh]
    \centering
    \includegraphics[width=\textwidth]{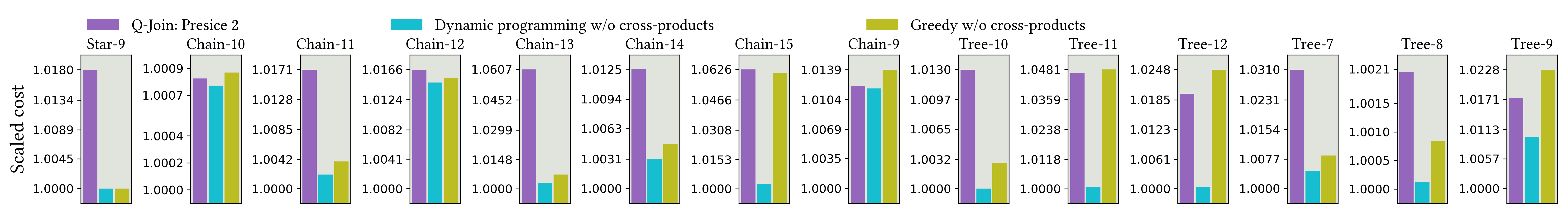}
    \caption{Precise 2 results using Gurobi solver}
    \label{fig:precise_2_gurobi_2}
    \Description[Precise 2 results using Gurobi solver]{}
\end{figure*}

Second, to evaluate the Gurobi solver, we scaled up the problem sizes remarkably from the Precise 1 method, although the experiments were performed on a standard laptop. The results are presented in Fig.~\ref{fig:precise_2_gurobi_1} and Fig.~\ref{fig:precise_2_gurobi_2}. We can see that finding the point that minimizes both cost and validity constraints becomes harder when the problem sizes increase. 

\begin{figure*}[tbh]
    \centering
    \includegraphics[width=\textwidth]{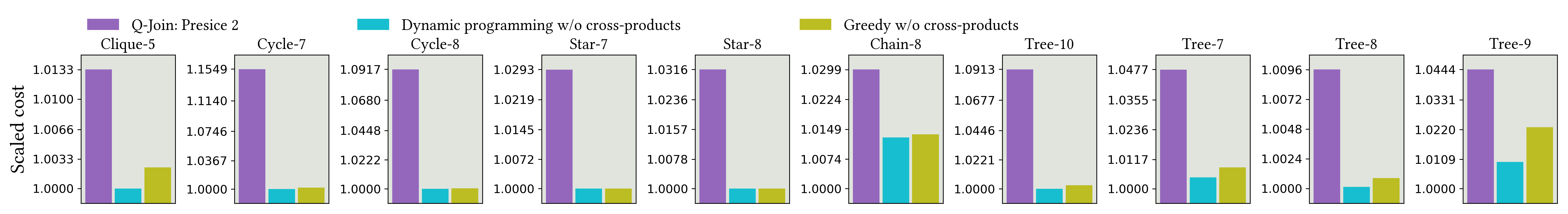}
    \caption{Precise 2 results using D-Wave's Leap Hybrid solver}
    \label{fig:precise_2_dwave_LeapHybridSampler}
    \Description[Precise 2 results using D-Wave's Leap Hybrid solver]{}
\end{figure*}

Slightly unexpectedly, the Leap Hybrid solver did not perform as well as we expected, as shown in Fig.~\ref{fig:precise_2_dwave_LeapHybridSampler}. The solver does not have tunable hyperparameters, which we would be able to adjust to obtain better results. On the other hand, we used the developer access to the solver, which is limited to only one minute of quantum computing access per month. Finally, we did not include the results from the D-wave quantum solver due to space limitations since the solver did not scale to these cases. 

\subsection{Evaluating heuristic formulation}

The key motivation behind the heuristic formulation is to tackle even larger query graphs. Our main goal is to demonstrate that this algorithm reaches acceptable results with superior scalability compared to the previous Precise 1 and 2 formulations. The results also indicate that Theorem \ref{thm:greedy_bound} is respected in practice. The optimal results are computed with dynamic programming without cross products. Due to space limitations, we only included the results from the Gurobi solver, which we consider the most demonstrative, and we had unlimited access to it since it runs locally. 

The results are presented so that we have computed and scaled the difference between each pair of methods. A value that differs from 0 indicates that the two methods gave different join trees with different costs. Since one of the methods is near-optimal (DP without cross products), it is clear which method produced the suboptimal result. This way, we can compare all three methods at the same time. In all cases, we can see that the heuristic algorithm respects Theorem \ref{thm:greedy_bound} very well in practice, so the difference between quantum and greedy is always $0$.

\begin{figure}
    \centering
    \includegraphics[width=\linewidth]{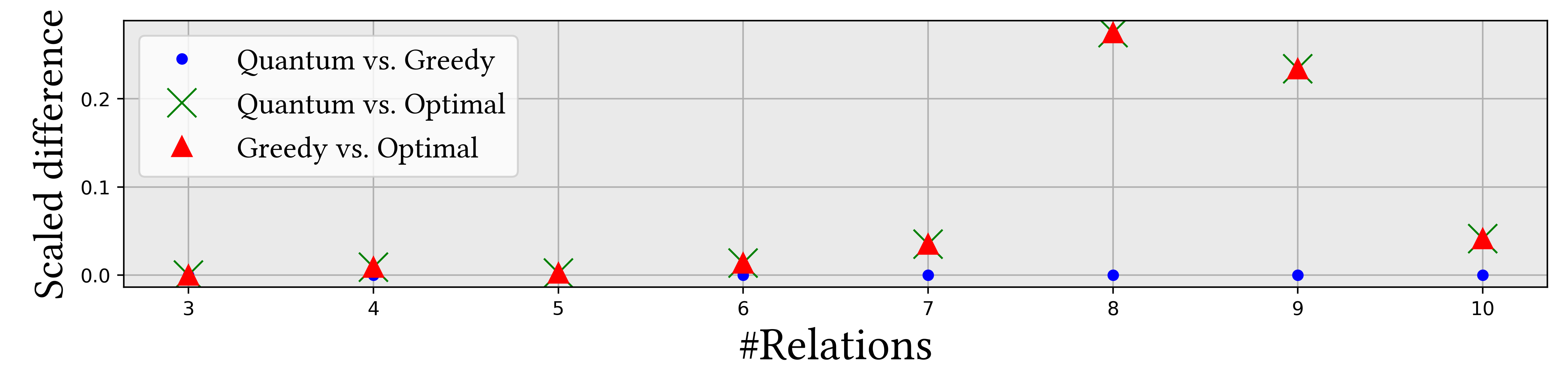}
    \caption{Heuristic results for clique query graphs using Gurobi solver}
    \label{fig:clique_accuracies}
\end{figure}

Fig.~\ref{fig:clique_accuracies} shows the results of applying the heuristic method to clique query graphs. Although these results are good, the scalability in this hard case is modest. On the other hand, we are unaware of any quantum computing research that would have outperformed this scalability in the case of clique graphs. For example, the most scalable method \cite{10.14778/3632093.3632112} excluded clique graphs from their results.

\begin{figure}
    \centering
    \includegraphics[width=\linewidth]{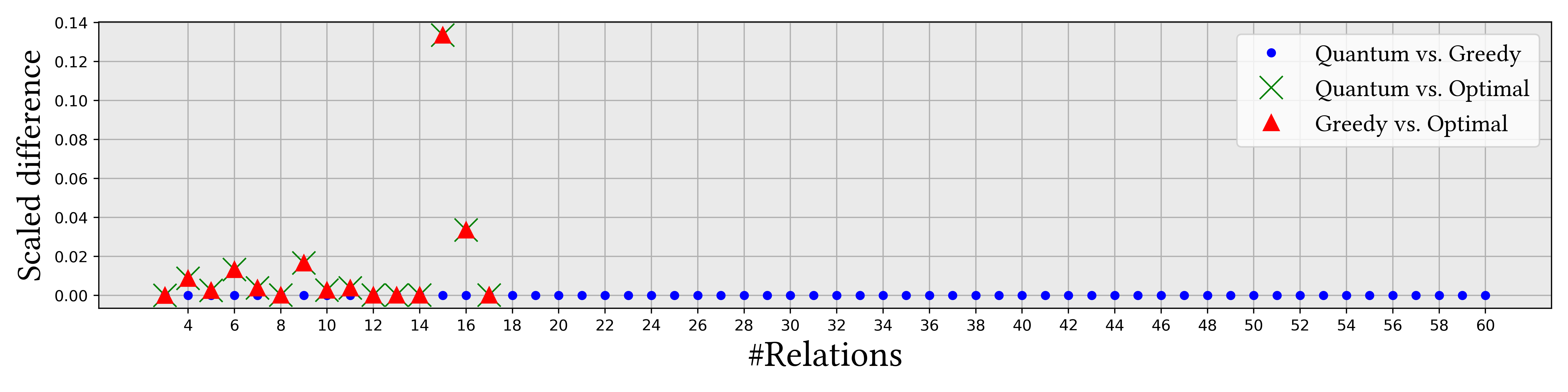}
    \caption{Heuristic results for tree query graphs using Gurobi solver}
    \label{fig:tree_accuracies}
\end{figure}

\begin{figure}
    \centering
    \includegraphics[width=\linewidth]{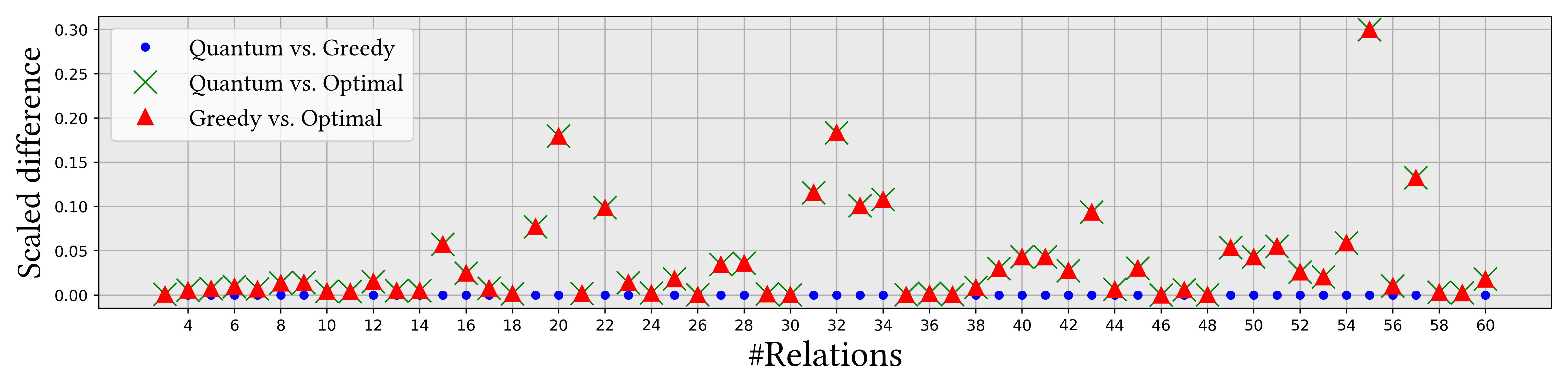}
    \caption{Heuristic results for chain query graphs using Gurobi solver}
    \label{fig:chain_accuracies}
\end{figure}

\begin{figure}
    \centering
    \includegraphics[width=\linewidth]{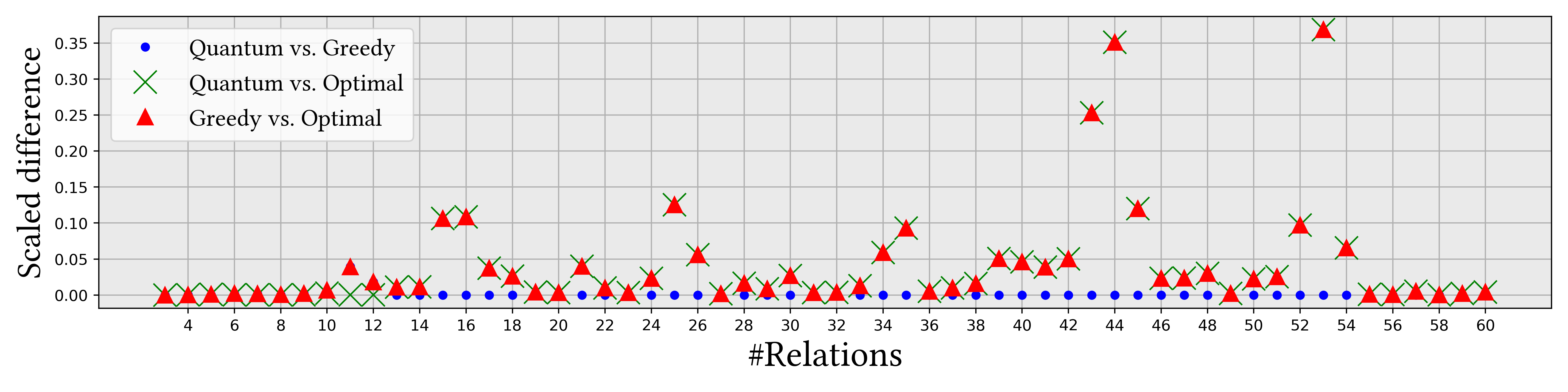}
    \caption{Heuristic results for cycle query graphs using Gurobi solver}
    \label{fig:cycle_accuracies}
\end{figure}

The results for the tree (Fig.~\ref{fig:tree_accuracies}), chain (Fig.~\ref{fig:chain_accuracies}), cycle (Fig.~\ref{fig:cycle_accuracies}), and star graphs demonstrate the best scalability. We computed the results up to 60 tables to demonstrate advantageous scalability over the most scalable method in the previous research \cite{10.14778/3632093.3632112} where they considered queries up to 50 relations.

We exclude the results for the star query graphs because, in this case, all three methods performed identically across up to 60 graphs and over 20 iterations, with no difference observed (a relative scaled difference of 0). These results may be because star graphs typically do not benefit from cross products \cite{10.14778/3632093.3632112}, and our method, which excludes them, performs better with such types of queries.

\section{Appendix: Comparison with quantum-inspired digital annealing}

\begin{figure}[!ht]
\centering
\begin{subfigure}{.5\textwidth}
  \centering
  \includegraphics[width=.95\linewidth]{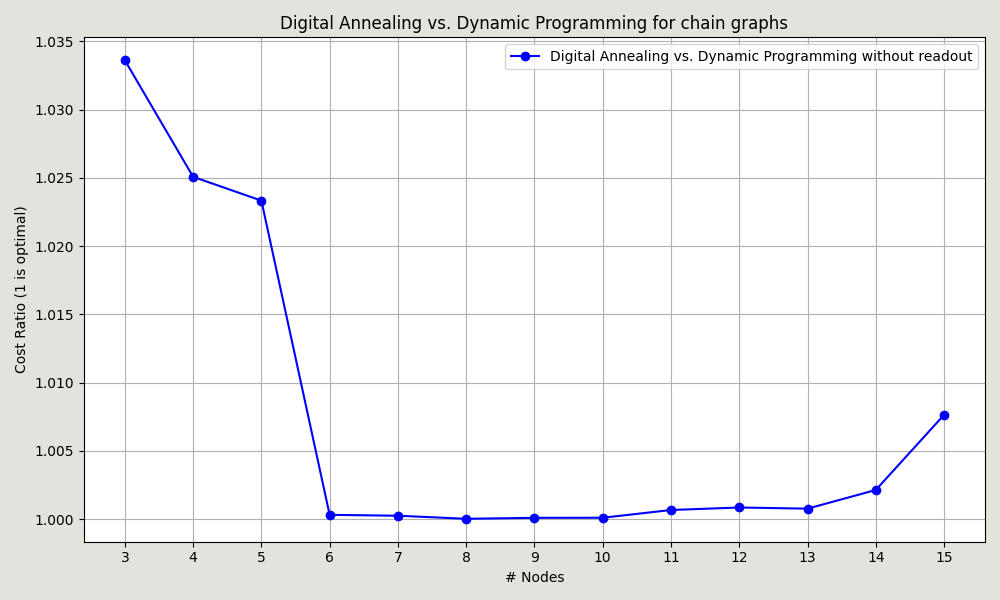}
  \caption{Chain graphs with varying number of nodes solved with QUBO formulation proposed in \cite{10.14778/3632093.3632112}}
  \label{fig:vldb24_chain}
\end{subfigure}%
\hfill
\begin{subfigure}{.5\textwidth}
  \centering
  \includegraphics[width=.95\linewidth]{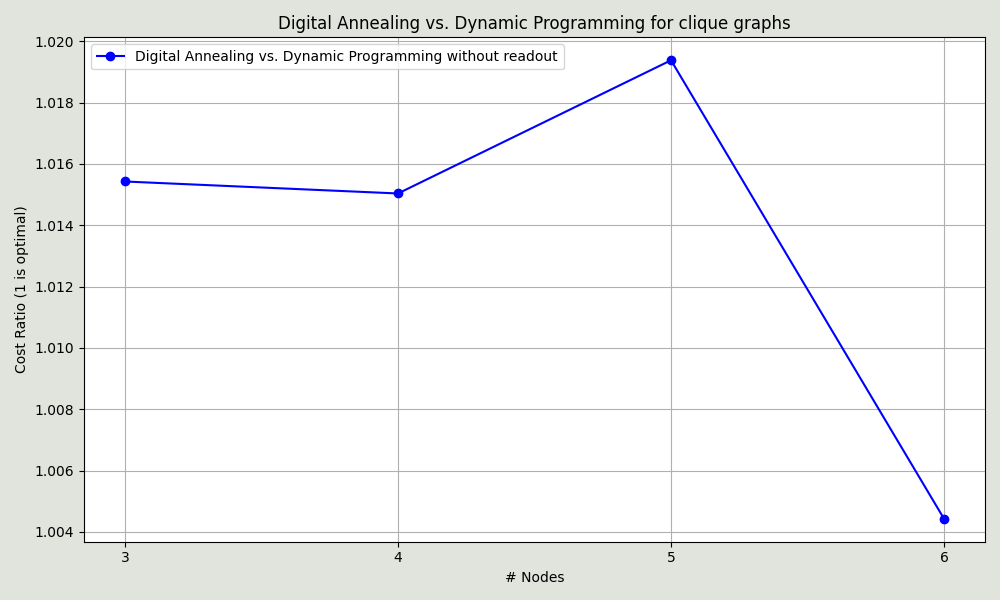}
  \caption{Clique query graphs with varying number of nodes solved with QUBO formulation proposed in \cite{10.14778/3632093.3632112}}
  \label{fig:vldb24_clique}
\end{subfigure}
\caption{A figure with two subfigures}
\label{fig:test}
\end{figure}

\begin{figure}[!ht]
\centering
\begin{subfigure}{.5\textwidth}
  \centering
  \includegraphics[width=.95\linewidth]{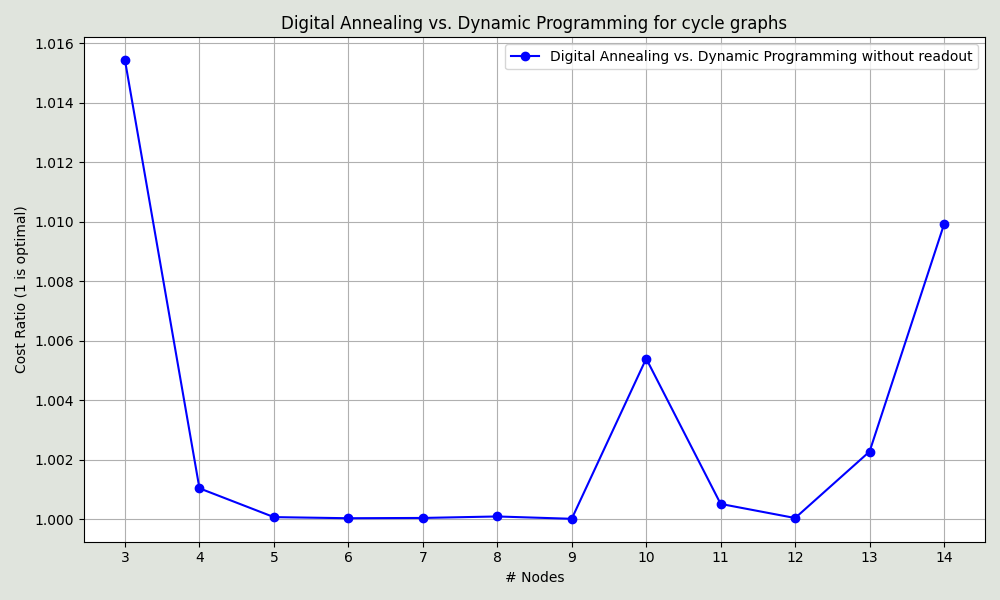}
  \caption{Cycle query graphs with varying number of nodes solved with QUBO formulation proposed in \cite{10.14778/3632093.3632112}}
  \label{fig:vldb24_cycle}
\end{subfigure}%
\hfill
\begin{subfigure}{.5\textwidth}
  \centering
  \includegraphics[width=.95\linewidth]{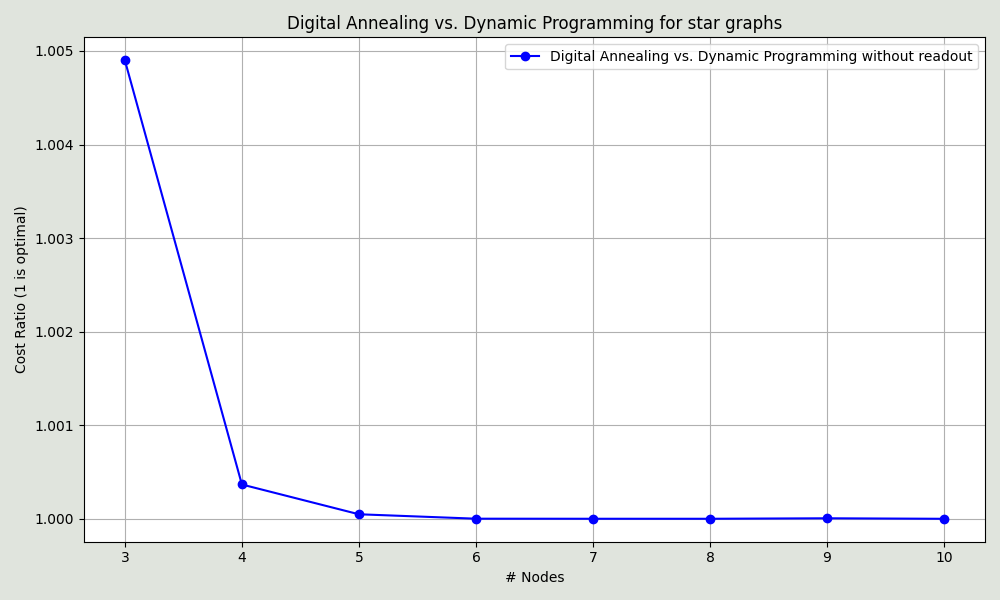}
  \caption{Star query graphs with varying number of nodes solved with QUBO formulation proposed in \cite{10.14778/3632093.3632112}}
  \label{fig:vldb24_star}
\end{subfigure}
\caption{A figure with two subfigures}
\label{fig:test}
\end{figure}

We evaluated the method proposed in \cite{10.14778/3632093.3632112} with the same workloads we used and present the results in Figures \ref{fig:vldb24_chain}, \ref{fig:vldb24_clique}, \ref{fig:vldb24_cycle}, and \ref{fig:vldb24_star}. The method is the improved algorithm from \cite{Schonberger_Scherzinger_Mauerer}. The authors propose a novel readout technique that improves the results. Since we could not access special quantum-inspired hardware, such as a digital annealer, we used the Gurobi solver, which returns only a single result by default. Thus, the readout technique was not applicable. 

Nevertheless, the results still demonstrate that the method reaches a comparable accuracy to ours, which is optimal or close to optimal. Their method seems to be able to identify beneficial cross products. We have computed the exact results with dynamic programming and compared relative cumulative costs between the methods. Due to higher-order terms in our method, which currently have to be rewritten into quadratic format, their method is still more scalable than ours. On the other hand, our theoretical bounds, the more straightforward variable definitions, and the novel usage of the higher-order model show specific improvements over their methods. We are also positive that our model admits features that make it easier to expand for outer joins and include more complex dependencies between the predicates.

\end{document}